\documentclass[fleqn,usenatbib]{mnras}

\usepackage{mathptmx}

\usepackage[T1]{fontenc}

\DeclareRobustCommand{\VAN}[3]{#2}
\let\VANthebibliography\thebibliography
\def\thebibliography{\DeclareRobustCommand{\VAN}[3]{##3}\VANthebibliography}

\usepackage{graphicx}	
\usepackage{amsmath}	
\usepackage{amssymb}	
\usepackage{url}
\usepackage{adjustbox}
\usepackage{caption}
\usepackage{subcaption}
\usepackage{mathtools}
\usepackage{amsfonts}
\usepackage{todonotes}
\usepackage[capitalize]{cleveref}
\usepackage{multirow}
\usepackage{array}
\usepackage{rotating}
\usepackage{hyperref}
\newcolumntype{H}{>{\setbox0=\hbox\bgroup}c<{\egroup}@{}}
\usepackage{ulem}

\title[Eddy profile and intrinsic temperature]{A physically derived eddy parameterization for giant planet atmospheres with application on hot-Jupiter atmospheres.}

\author[A. Arfaux et al.]{
Anthony Arfaux,$^{1}$\thanks{E-mail: anthony.arfaux@univ-reims.fr}
Panayotis Lavvas,$^{1}$
\\
$^{1}$Groupe de spectrom\'erie mol\'eculaire et atmosph\'erique, Universit\'e de Reims Champagne Ardenne, Reims, France\\
}

\date{Accepted 2023 April 7. Received 2023 April 6; in original form 2023 January 13}

\pubyear{2022}

\begin{document}
\label{firstpage}
\pagerange{\pageref{firstpage}--\pageref{lastpage}}
\maketitle

\begin{abstract}
We present a parameterization for the eddy diffusion profile of gas giant exoplanets based on physical phenomena and we explore how the parameterized eddy profile impacts the chemical composition, the thermal structure, the haze microphysics, and the transit spectra of 8 hot-Jupiters. Our eddy parameterization depends on the planetary intrinsic temperature (T$_{int}$ ), we thus evaluate how the increase of this parameter to values higher than those typically used ($\sim$100K) impacts the atmospheric structure and composition. Our investigation demonstrates that despite the strong impact of T$_{int}$ on the chemical composition of the deep atmosphere, the upper atmosphere is not affected for T$_{eq}$ $>$ 1300 K owing to high altitude quench levels at these conditions. Below this threshold, however, the larger atmospheric temperatures produced by increasing T$_{int}$ affect the quenched chemical composition. Our parameterization depends on two parameters, the eddy magnitude at the radiative-convective boundary (K$_0$) and the corresponding magnitude at the homopause (K$_{top}$). We demonstrate that, when using common K$_0$ and K$_{top}$ values among most of the different planet cases studied, we derive transit spectra consistent with Hubble Space Telescope observations. Moreover, our simulations show that increasing the eddy profile enhances the photochemical production of haze particles and reduces their average radius, thus providing a steeper UV-Visible slope. Finally, we demonstrate for WASP-39b that the James Webb Space Telescope observations provide improved constraints for the hazes and clouds and we show that both components seem necessary to interpret the combined transit spectrum from HST and JWST observations. 
\end{abstract}



\section{INTRODUCTION}

Atmospheric simulations can be conducted in one, two or three dimensions.
The choice depends on the processes studied and how much their dimensions can be reduced. 
Despite the wealth of information they can provide, 3D models require powerful computation facilities and long execution times.
Therefore 3D simulations are limited in the number and/or complexity of the processes they can simulate, for instance using reduced chemical schemes or neglecting the complex microphysics of particles \citep{Venot19,Steinrueck21,Tsai22b}.
The use of a 1D model can overcome these limitations, however, it cannot reproduce fundamentally 3D processes like atmospheric dynamics.
Nevertheless, when working on disequilibrium chemistry and haze microphysics via a 1D model, vertical atmospheric mixing has to be taken into account \citep{Spiegel09,Moses11,Lavvas14,Lavvas17,Wong20,Gao20,Kawashima21}.
This vertical transport results from multiple phenomena at different scales such as convection \citep{Gierasch85,Ackerman01}, turbulent flow arising from horizontal shear stress \citep{Holton04,Steinrueck20}, and gravity waves \citep{Lindzen71,Hunten75,Chamberlain78}. 
Most of these contributions happen in a 3D framework, thus 1D models cannot generate them.
We further note that some of these processes, like shear stress and wave breaking, happen on small enough scales to be difficult to fully capture even by typical 3D models.

A common way to account for vertical transport is via a diffusion-like process, so called eddy diffusion or eddy mixing (K$_{zz}$), providing a vertical flux opposed to the particle density gradient \citep{Colegrove65}.
Multiple ways have been used in the past decades to determine the eddy coefficient.
$K_{zz}$ can be treated as a free-parameter and derived from composition observations of chemically inert species \citep{Spiegel09,Wong20,Kawashima21}.
This approach is usually sensitive to an assumed constant eddy profile.
The eddy profile may also be derived from Global Circulation Models (GCMs).
In such circumstances, a first approximation is based on the root mean square (rms) vertical velocities \citep{Moses11,Smith98}:
$	K_{zz} = w_z L_z	$,
where $w_z$ is the horizontally averaged global rms vertical velocity and $L_z$ a length scale related to the phenomenon.
A common approximation for the length scale is the scale height, but this may overestimate the eddy profile and a fraction ($\sim$1/10th) of the scale height may be more reliable \citep{Smith98}.
The eddy profile can also be derived based on passive tracers distribution from the 3D models \citep{Parmentier13,Zhang18a,Zhang18b,Steinrueck21,Komacek22a}.
However, 3D GCM calculations do not take into account sub-grid scale effects, thus may underestimate the strength of eddy diffusion.
Moreover this method requires calculations specific to each planet which are currently limited to a few planetary cases.
Therefore, it would be beneficial to have a parameterization for the eddy mixing profile based on physical arguments that could be readily applied to different atmospheric environments.
\cite{Zhang18a,Zhang18b} developed a parameterization for the eddy diffusivity based on the long range atmospheric circulation. However, their parameterization depends on inputs from GCMs, like the vertical wind speed, the dayside-nightside temperature difference profile, and characteristic times for various processes (drag, chemistry, radiation, etc). Although these can be estimated, application of their parameterization suggests better results when the required input is based on GCM results \citep{Komacek19}.
Instead we want to develop a scalable parameterization with a minimal number of free parameters, which could be adjusted to GCM /observational constraints. We compare our parameterization to the \cite{Zhang18a} and \cite{Komacek19} results further below.

In the deep atmosphere, the eddy profile is dominated by convection, and therefore it is affected by the heating mechanisms occurring in the planet interior.
Planet interiors reach high temperatures during planet formation before cooling down during the long term planetary evolution.
This cooling can be delayed or compensated by other phenomena like radioactivity.
For close-in gaseous giants, the observed mass-radius relationship implies much hotter interior temperature than those anticipated by planet formation models to explain the planet inflation \citep{Bodenheimer01,Guillot02}.
Indeed, many hot-Jupiters present radii larger than what their mass suggests implying the existence of mechanisms depositing energy in the planet interior, below the radiative-convective boundary (RCB).
Owing to their short orbital distance, tidal forces exerted by the host star on the planet can be responsible for the heating mechanism leading to the planet inflation \citep{Bodenheimer01,Winn05}.
Another proposition by \cite{Guillot02} suggest that this excess energy can be deposited by convective instabilities, while \cite{Batygin10} suggested the role of ohmic dissipation, whereby the interaction between the ions in the atmosphere and the planetary magnetic field creates currents flowing down to deep regions where they dissipate their energy and heat the planet interior.
Usually, the thermal flux arising from the planet interior to the atmosphere is accounted via a simple black body emission coming from below the radiative-convective boundary, with a temperature $T_{int}$ called intrinsic temperature.
Based on the understanding of planetary interiors from the solar system giant planets, typical values for $T_{int}$ are around 100K.
However, \cite{Thorngren19b} proposed values far hotter up to 700 K for hot-Jupiters based on the observed planet mass-radius relationship, which matches results from the ohmic dissipation mechanism \citep{Thorngren18}.
Apart from affecting the eddy mixing, such high intrinsic temperatures could impose modifications on the thermal structure and chemical composition of the atmosphere \citep{Fortney20}.

In this study, we develop a parameterization for the eddy profile of gas-giant atmospheres based on the main physical phenomena expected to drive their atmospheric mixing and we explore the impact of the high T$_{int}$ values on the parameterized eddy profile. We further investigate how the parameterized eddy profiles affect the chemical composition and thermal structure of hot-Jupiter atmospheres, as well as the implications of these modifications on the haze distribution of such planets. 
We focus on 8 planets we have previously studied \citep[][]{Arfaux22} in view of their transit observations with the Hubble Space Telescope (HST), as well as, compare with novel results from the James Webb Space Telescope (JWST). 
In \cref{Sec:Method}, we provide a description of our 1D model (\cref{SSec:Model}), the intrinsic temperatures used (\cref{SSec:Tint}) and our eddy profile parameterization (\cref{SSec:Eddy}) and we dwell on the specific case of WASP-39b in light of the recent JWST observations (\cref{SSec:W39Cloud}).
Our results are described in \cref{Sec:Results}, which is separated into two main sections, the first focusing on the intrinsic temperature (\cref{SSec:TintRes}) and the second on the eddy profile parameterization (\cref{SSec:EddyRes}) .
In \cref{Sec:Discussion}, we discuss various implications of our results and \cref{Sec:Conclusions} presents our final conclusions.

\section{METHOD}
\label{Sec:Method}

Here we a provide a brief description of the model used and we detail the parameterization of the eddy profile and the intrinsic temperature.
We also explore new possibilities for WASP-39b in the light of new observations conducted with the James Webb Space Telescope.

\subsection{Model description}
\label{SSec:Model}

We use a self-consistent 1D model to simulate exoplanet atmospheres and derive theoretical transit spectra.
This model couples disequilibrium chemistry, haze microphysics and radiative transfer to obtain a detailed view of the interactions between these three components.
The model is further detailed in \cite{Arfaux22} and we provide here the main points only.
The disequilibrium chemistry model calculates the chemical composition profiles in the atmosphere taking into account their transport, driven by molecular and eddy diffusion, and the (photo)chemical reactions of the different species included.
In the deep atmosphere (100 to 1000 bar), the high temperatures and pressures keep the atmosphere close to thermochemical equilibrium.
At higher altitudes (lower pressures), first transport and then photolysis reactions start to modify the composition away from the thermochemical equilibrium at rates different for each species.
The haze microphysics model calculates the haze size and density distributions accounting for (spherical) particle coagulation, sublimation, settling and transport via eddy diffusion.
The particles are produced around 1 $\mu$bar and they settle towards the deep atmosphere where they sublimate by the high local temperatures.
A strong gravity or eddy diffusion enhances the particle transport and impedes their coagulation, thus leading to smaller and more numerous particles.
The radiative transfer model computes the radiation field and thermal structure of the atmosphere taking into account the disequilibrium chemical composition and the haze distribution. 
Convection dominates the deep atmosphere thermal structure while the upper atmosphere is assumed in radiative equilibrium.
The location of the transition from the convective regime to the radiative regime is mainly impacted by the intrinsic flux outgoing the planet interior.

These three models are coupled, thus allowing to account for the different feedbacks taking place among them.
For example, the composition profiles of radiatively active species can dramatically modify the radiation field and the thermal structure, while the presence of an homopause in the upper atmosphere results in a very hot thermosphere.
Moreover, photolysis reactions inform on the mass flux produced by the photochemistry of haze precursors, thus on the efficiency of haze formation.
Hazes present in the upper atmosphere absorb stellar UV radiation and release this excess energy via conduction in the atmosphere, leading to larger local temperatures, and colder deep atmosphere temperatures.
These changes in the temperature profile modify the chemical composition and haze distribution, notably leading to larger particles as the hotter temperatures enhance the coagulation through Brownian motion.
These modifications therefore affect the radiation field, the temperature profile and the chemical composition.
All these complex feedbacks are self-consistently accounted in our coupled model.

In our previous study \citep{Arfaux22} we used the above model to retrieve constraints for the haze mass flux of different planets based on HST transit observations. The results from that study are summarized in \cref{Tab:Ini} and are used here as our initial scenarios. 
These results were obtained assuming a nominal eddy profile and a 100 K intrinsic temperature. The nominal eddy profile was derived based on the eddy diffusion used by \cite{Moses11} for HD-189733b but downscaled by a factor of 100 following more recent estimates of the eddy mixing efficiency \citep{Parmentier13}. However, in the specific case of HD-189733b, we chose to keep as nominal the profile used by \cite{Moses11} without a scaling factor, which provided a better fit to the observations \citep{Lavvas17,Arfaux22}.
Only for the case of WASP-39b, we conducted additional simulations that are described below (\cref{SSec:W39Cloud}) and we use these new parameters as initial scenario.

\begin{table}
\caption{Haze mass fluxes, production altitudes and metallicities retrieved in our previous work \protect\citep[][except for WASP-39b*]{Arfaux22} by fitting HST observations assuming the nominal eddy profile and T$_{Int}$ = 100 K intrinsic temperature. The haze production altitude and metallicities are kept constant in the current study (except for WASP-39b for which new results indicate a higher metallicity).}
\label{Tab:Ini}
\centering
\begin{tabular}{c|ccc}
Planet   	 	&	Haze mass flux		&	Haze production 	&	Metallicity 		\\
			&	($g.cm^{-2}.s^{-1}$)	&	altitude ($\mu$bar)	&	($\times$solar)	\\
\hline
HAT-P-12b       	& 	 1.0e-14			&	1				&	1			\\
WASP-39b*      	& 	 1.0e-15			&	1				&	10			\\
HD-189733b     & 	 9.0e-12			&	0.1				&	1			\\
WASP-6b      	& 	 1.0e-14			&	1				&	0.1			\\
HAT-P-1b      	& 	 7.0e-16			&	1				&	1			\\
HD-209458b     & 	 3.3e-15			&	1				&	1			\\
WASP-31b      	& 	 1.8e-17			&	10				&	0.1			\\
WASP-17b     	& 	 $<$1e-16			&	1				&	1			\\
\end{tabular}
\end{table}

\begin{figure}
\includegraphics[width=0.5\textwidth]{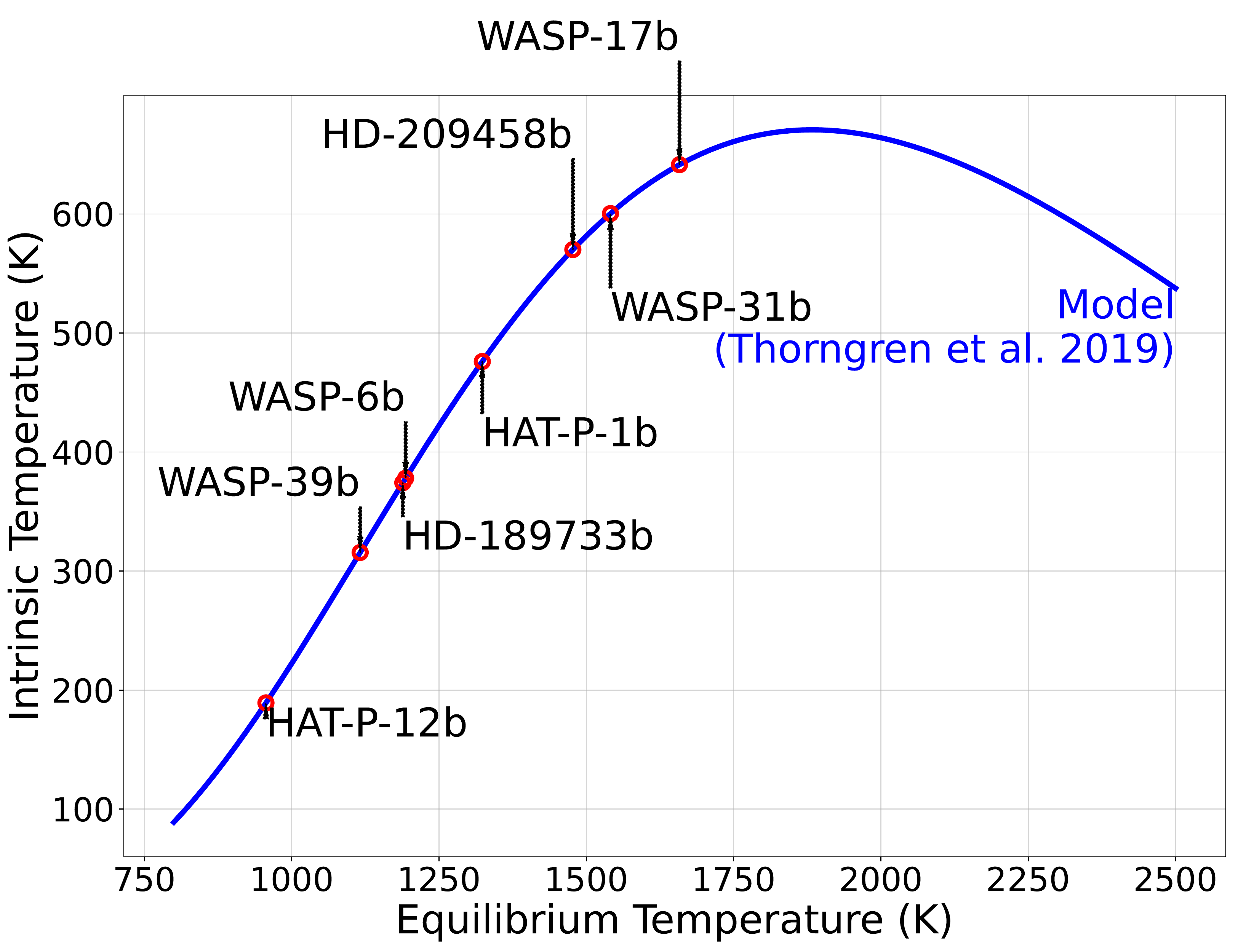}
\caption{Intrinsic temperature of the studied planets. The blue line corresponds to the model provided by \protect\cite{Thorngren19b} and the red circles represent the 8 planets of the current study.}
\label{Fig:Tint}
\end{figure}

\subsection{Intrinsic temperature}
\label{SSec:Tint}

In our previous work \citep{Arfaux22}, we used a value of 100 K for the intrinsic temperature, classically used for solar system gas giant planets.
However, \cite{Thorngren19b} suggest much hotter values for the intrinsic temperature based on the observation of 281 gas giants \citep{Thorngren18}.
They provide an equation to calculate the intrinsic temperature assuming equilibrium between the stellar flux and the outgoing intrinsic flux, with a parameterization expressed as a function of the planet equilibrium temperature.
The intrinsic temperature also depends on the system mass and age, however, in our purpose of comparison with the previously used T$_{int}$ = 100 K, we consider the \cite{Thorngren19b} relationship as a sufficient approximation.
In addition, past $\sim$0.1 Gyr, the system age is assumed to demonstrate no effects on the interior structure \citep{Thorngren18}, which corresponds to the case of all the planets included in our study, while  \cite{Thorngren19b} argue that the system mass add at most a modest uncertainty to their T$_{int}$ estimation.
The calculated equilibrium and intrinsic temperatures of the planet we are studying are presented in \cref{Fig:Tint}.
Our sample of planets presents a large range of intrinsic temperatures from 189 K for HAT-P-12b, the coldest planet in this study, to 641 K for WASP-17b.
Within our sample, the intrinsic temperature presents a monotonic behavior increasing with the planet equilibrium temperature.
We must however note that for hotter planets the intrinsic temperature decreases with increasing equilibrium temperatures.

\subsection{Eddy profile}
\label{SSec:Eddy}

In this section, we present our eddy profile parameterization suitable for all gas giants exoplanets.
In a first part, we discuss the theoretical aspects of our parameterization by presenting the main processes driving the strength of the eddy diffusion (\cref{SSEC:EddyTh}).
In a second part, we provide the mathematical details of our parameterization (\cref{SSEC:EddyPara}).

\subsubsection{Theory}
\label{SSEC:EddyTh}

We derive our eddy description by first separating the atmosphere into three distinct regions, where the phenomena driving the strength of the eddy diffusion are different. 
First, below the radiative-convective boundary (RCB), determined with the Schwarzschild's criterion, the vertical transport is driven by convection.
In this regime, the eddy profile is estimated by the eddy diffusion for heat transfer \citep{Gierasch85,Ackerman01}:
\begin{equation}
	K_{zz} = \frac{H}{3} \left( \frac{L}{H} \right)^{4/3} \left( \frac{R F}{\mu \rho c_p} \right)^{1/3}
\end{equation}
where $H$ is the scale height, $R$ the ideal gas constant, $\mu$, $\rho$ and $c_p$ the atmospheric mean molecular weight, mass density and heat capacity, respectively.
$F$ is the convective flux calculated as $F=\sigma T_{eff}^4$, with $T_{eff}$ the planet equilibrium temperature.
$L$ is the mixing length, assumed to be a fraction of the scale height, the scaling factor being provided by the ratio between the atmospheric and adiabatic lapse rates.
A minimum value $\Lambda = 0.1$ is also defined for this factor, providing: $L = H \times max(\Lambda, \frac{\Gamma}{\Gamma_{ad}})$.

Above the convective boundary, horizontal shear stress and gravity wave break down are the main sources of vertical transport.
This region is currently poorly described theoretically in terms of eddy diffusion.
Multiple works attempt to provide a theoretical framework for this region modeling the diffusion through random gravity wave break down \citep{Lindzen71,Hunten75}: large scale atmospheric flow is perturbed by gravity waves that locally transport material up and down.
On average, there is no global transport of species and particules, unless the gravity waves collapse.
If this breaking happens after a time $t$ different from a multiple of the half-oscillation period of the wave ($t \ne n \times T/2$ with $T$ the wave period and $n$ an integer), material will be definitely carried to another region of the atmosphere, resulting in a non-zero net flux.
The eddy magnitude is increasing over this region starting from a minimum value above the convective boundary and approaches a constant value at the homopause.
This increase is assumed to follow a power law:
\begin{equation}
	K_{zz} \propto \rho^{-1/2}
\end{equation}
Indeed, according to \cite{Chamberlain78}, as the wave energy evolves as the square of its amplitude, the eddy magnitude has to be proportional to the invert square of the density for the energy conservation.
In addition, this 0.5 value matches the behavior expected from GCM simulations for the eddy profile in the radiative region of the atmosphere \citep{Parmentier13}, though \cite{Agundez14} and \cite{Steinrueck21} found slightly different values (0.65 and 0.9, respectively).

Finally, above the homopause, that is the layer in the atmosphere where molecular diffusion takes over eddy diffusion, the eddy profile can be assumed constant.
\cite{Koskinen10} proposed to scale the homopause eddy coefficient of Jupiter to derive the homopause value of hot-Jupiters:
\begin{equation}
	K_{zz} = K_{zz}^J \frac{v}{v_J} \frac{g_J}{g} \frac{T}{T_J}
\end{equation}
where $v$ and $v_J$ are the characteristic turbulent velocities, $g$ and $g_J$ the gravities and $T$ and $T_J$ the equilibrium temperatures for the studied planet and Jupiter, respectively, and 
$K_{zz}^J$ = 10$^{7} cm^2.s^{-1}$ is the homopause eddy coefficient of Jupiter.

\subsubsection{Parameterization}
\label{SSEC:EddyPara}

\begin{figure}
\includegraphics[width=0.5\textwidth]{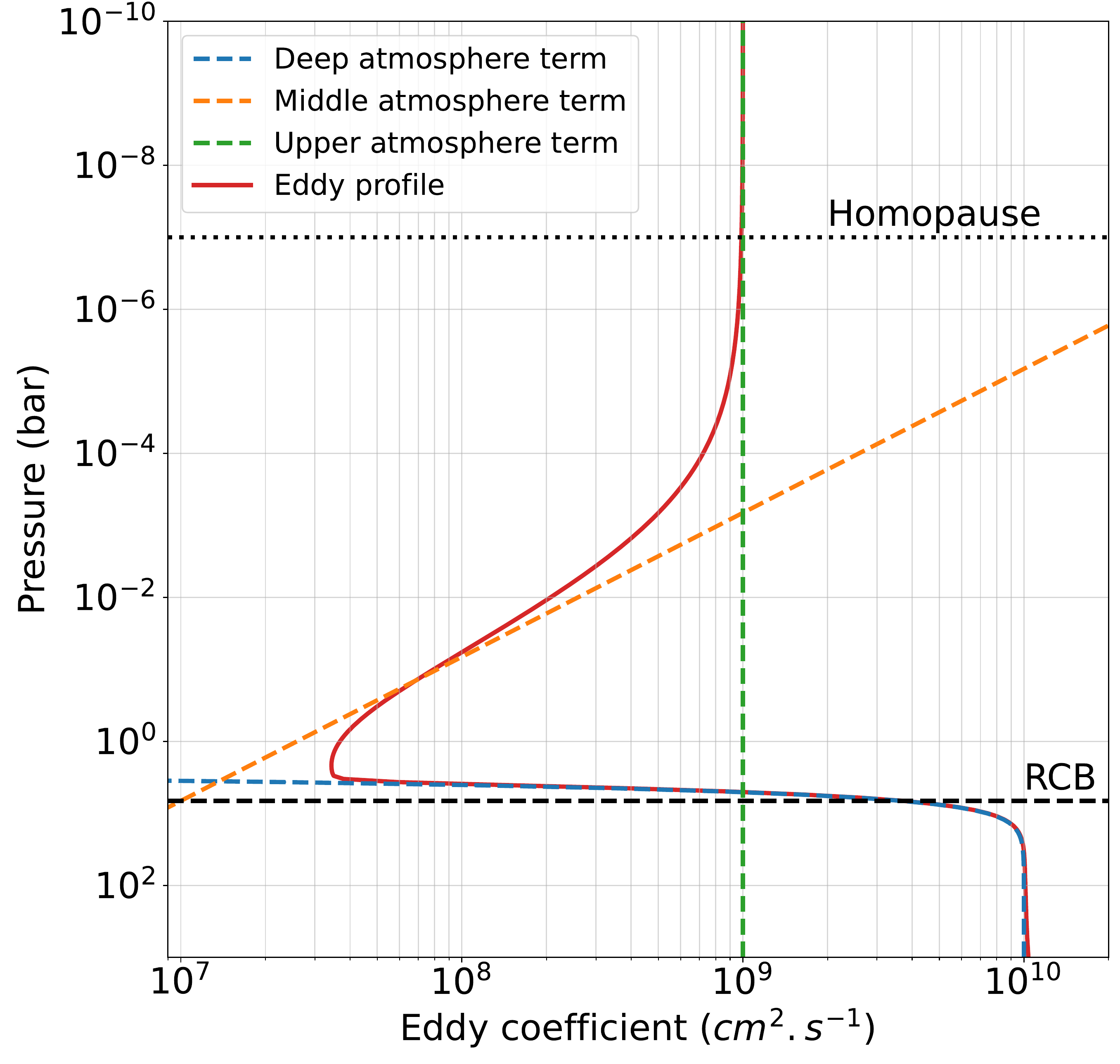}
\caption{Illustration plot of the eddy parameterization and its different components. The temperature profile is assumed isothermal with T = 1000 K, the mean molecular weight is constant at 2.3 g/mol. The RCB is randomly assumed around the 6 bar altitude. $K_{bot}$ and $K_{top}$ are randomly set to 10$^{10}$ and 10$^9 cm^2.s^{-1}$, respectively.}
\label{Fig:DemoEddy}
\end{figure}

Our parameterization is based on these three different regions of the atmosphere.
In the deep convective region, the eddy profile is assumed to follow an exponential that equals a fixed value $K_{bot}$ at the bottom boundary ($P=P_{bot}$) and approaches zero above the convective boundary ($P=P_{RCB}$), thus providing (blue dashed line in \cref{Fig:DemoEddy}):
\begin{equation}
	\label{eq:deep}
	K_{zz}^{Deep}(P) = K_{bot} exp\left[-P_{RCB}  \left( \frac{1}{P} - \frac{1}{P_{bot}} \right) \right]
\end{equation}
The value of $K_{bot}$ is calculated based on the heat convective diffusion coefficient from \cite{Ackerman01} setting the minimum scaling factor $\Lambda$ to 0.1 as they recommended.

From the convective boundary to the top of the atmosphere, we use an interpolation formula of the form:
\begin{equation}
	K_{zz}^{Mid}(\rho) = \frac{U(\rho) L(\rho)}{U(\rho) + L(\rho)}
\end{equation}
with $K(\rho) = U(\rho)$ when the density is small (upper boundary) and $K(\rho) = L(\rho)$ at larger densities (lower boundary).
We chose $U(\rho) = K_{top}$ as a constant in the upper atmosphere (green dashed line in \cref{Fig:DemoEddy}), corresponding to the homopause value obtain based on \cite{Koskinen10} scaling of Jupiter's eddy diffusion coefficient.
The characteristic turbulent velocity is unknown and we keep the scaling $\frac{v}{v_J} = 5$ proposed for HD-209458b.

In the middle atmosphere, the eddy profile is assumed to follow a power law of the shape:
$$ L(\rho) = \alpha \rho^{-\gamma} $$
Assuming $L(\rho_0) = K_0$, we obtain (orange dashed line in \cref{Fig:DemoEddy}):
$$ L(\rho) = K_0 \left( \frac{\rho_0}{\rho} \right)^{\gamma} $$
with $K_0$ the minimum eddy coefficient.
We chose $\gamma=0.5$ based on \cite{Chamberlain78}.
$\rho_0$ is chosen as the convective boundary and the eddy diffusion value at this altitude is set to $10^7 cm^2.s^{-1}$.
We therefore obtain:
\begin{equation}
	\label{eq:top}
	K_{zz}^{Mid}(\rho) = \frac{K_0 K_{top} \rho_{RCB}^{\gamma}}{K_0 \rho_{RCB}^{\gamma} + K_{top} \rho^{\gamma}}
\end{equation}
This is similar to the the parameterization proposed by \cite{Vuitton19} used for Titan’s atmosphere, albeit their calculation assumes different parameters and is based on the pressures.

Putting \cref{eq:deep,eq:top} together leads to:
\begin{equation}
	K_{zz} = \frac{K_0 K_{top} \rho_{RCB}^{\gamma}}{K_0 \rho_{RCB}^{\gamma} + K_{top} \rho^{\gamma}}
	+ K_{bot} exp\left[-P_{RCB}  \left( \frac{1}{P} - \frac{1}{P_{bot}} \right) \right]
\end{equation}

To avoid a two sharp transition at the RCB from the deep atmosphere drop led by $K_{zz}^{Deep}(P)$ to the middle atmosphere increase driven by $L(\rho)$, we added a term $ 3 K_0 \left(\frac{\rho_0}{\rho}\right)^{-\gamma}$. This term allows to smooth the transition above the RCB without impacting the rest of the atmosphere in order to avoid numerical instabilities. 
 \cref{Fig:DemoEddy} presents an example of our parameterized eddy, as well as, the contributions of its different components for typical values of K$_{bot}$ and K$_{top}$. The major advantages of using such an estimate of the eddy profile are that it is driven by physical arguments and that it depends on 2 free parameters, which are the RCB ($K_0$) and homopause ($K_{top}$) eddy diffusivities, the latter depending on the characteristic turbulent velocity. We consider the other parameters to be sufficiently constrained and assume them to be fixed. Therefore, our parameterization allows for a quantitative investigation of the impact of eddy mixing on the observable parameters.

\subsection{The special case of Wasp-39b}
\label{SSec:W39Cloud}

\begin{figure*}
\includegraphics[width=\textwidth]{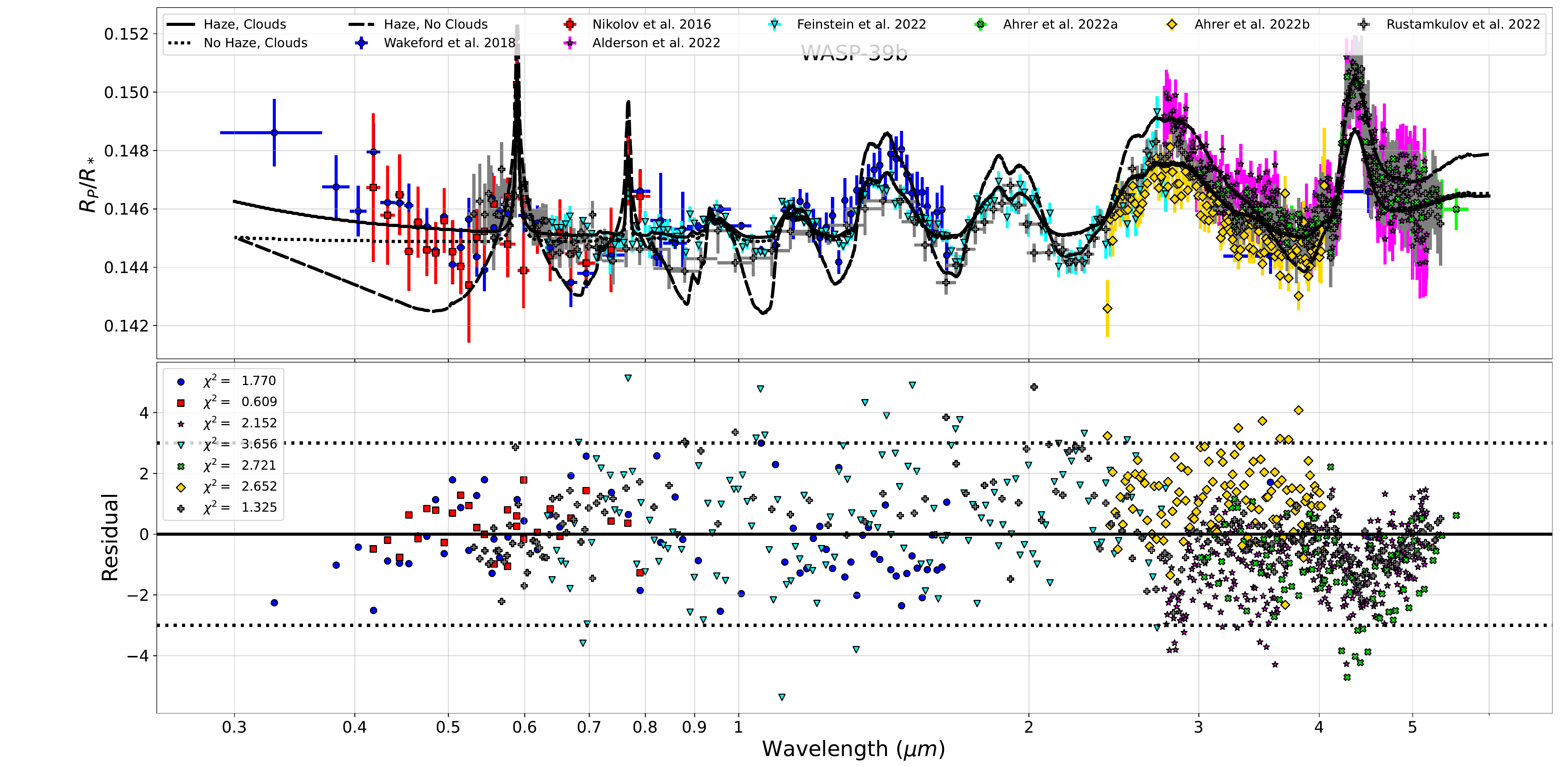}
\caption{Best fit (10$^{-15} g.cm^{-2}.s^{-1}$ haze mass flux and 10$\times$solar metallicity). 
Top panel: transit spectrum for haze \& cloud case (solid line), haze only case (dotted line) and cloud only case (dashed line). 
Bottom panel: residuals for WASP-39b haze \& cloud case. 
The transit spectra are numerically smoothed with a savgol filter to allow a better comparison with observations.}
\label{Fig:W39best}
\end{figure*}

As we move in the era of JWST observations our understanding of exoplanetary atmospheres will improve and the constraints we can set in the haze parameters are likely to be modified from what we got based on the HST observations. This is clearly demonstrated for the case of WASP-39b for which the first JWST results suggest an atmosphere of $\sim$10xsolar metallicity \citep{JWST22,Alderson22b,Ahrer22,Feinstein22,Rustamkulov22}. Including these novel constraints in our analysis has major ramifications for the haze mass flux required to match the UV HST observations while being consistent with the atmospheric metallicity inferred from the JWST observations. 

Our simulation with a higher (10$\times$solar) metallicity (using the high T$_{int}$) leads to lower transit depths in the UV-visible compared to solar metallicity case therefore indicates the presence of haze/clouds.
Considering a photochemical haze component with a mass flux of 2x10$^{-15} g.cm^{-2}.s^{-1}$ provides a good fit of the HST \citep{Wakeford18} and VLT observations \citep{Nikolov16}), while being consistent with the JWST/NIRSpec G395H observations \citep{Alderson22b}.
However, the resulting transit spectrum underestimates the observed transit depth of the JWST/NIRSpec PRISM \citep{Rustamkulov22}, the JWST/NIRISS \citep{Feinstein22} and the JWST/NIRCam \citep{JWST22,Ahrer22} observations at the minima between the molecular features at wavelengths larger than 0.8 µm.
This difference could imply that these spectral regions are affected by the presence of clouds.

To evaluate if clouds could replace the role of photochemical hazes in our fits, we evaluated transit spectra assuming only cloud particles. Our goal is to see if the clouds could mimic the UV-Visible spectral slope thus we considered a lognormal size distribution of spherical cloud particles centered at $<r>$ = 1 µm, which is at the lower range limit of typical cloud particles sizes \citep{Lee15,Ohno18,Gao18}.
For the number density we assumed 1 part.cm$^{-3}$, corresponding to the haze particle density at 10 mbar, and constant with altitude, expanding from 100 to 1 mbar.
Considering aerosols serving as nucleation sites for cloud, this assumption supposes that all the haze particles at 10 mbar are coated with the condensible species.
We also tested the impact of a cloud density decreasing with the atmospheric scale height with a particle density at 1, 0.1, 0.01 and 0.001 part.cm$^{-3}$ at the cloud base located at 0.1 bar.
This exploration did not yield a better fit than the constant cloud case.
Na$_2$S, MnS and MgSiO$_3$ are the most likely condensate candidates \citep{JWST22,Alderson22b,Feinstein22}.
Given that Na has a higher elemental abundance than Mn \citep{Lodders10} and that MgSiO$_3$ forms at large pressures preventing it to expand up to the probed region, we considered a Na$_2$S composition for the cloud optical properties \citep{Batalha20}.
The cloud only case produces a good fit of all available data, remaining almost within the 3$\sigma$ of every observations.
However, we note that the resulting spectrum (dashed line in \cref{Fig:W39best}) in the UV-visible range is much flatter compared to the HST and VLT data, the latter showing a strong slope.

We thus explored also the combined case of haze \& clouds. We evaluated spectra with different photochemical haze mass fluxes ranging from 5x10$^{-16}$ to 10$^{-14} g.cm^{-2}.s^{-1}$.
Despite that \cite{Ahrer22} and \cite{Rustamkulov22} are better fitted with the cloud only case based on the $\chi^2$, \cite{Alderson22b} and \cite{Feinstein22} tend to favor the haze \& cloud case.
We further note that \cite{Rustamkulov22} observations are in weak agreement with \cite{Feinstein22} in the  0.9 - 1.5 µm wavelength range, corresponding to the region where \cite{Rustamkulov22} reported saturation of the detector. 
We therefore consider the \cite{Feinstein22} observations to be more reliable at those wavelengths. 
In addition, \cite{Ahrer22} observations cover the wavelength range beyond 2 µm and are therefore weakly sensitive to haze absorption.
The precision of the HST \citep{Wakeford18}, VLT \citep{Nikolov16} and JWST/NIRSpec PRISM \cite{JWST22} observations do not allow to discriminate between the cloud only and haze \& cloud cases, providing similar $\chi^2$ with a relative difference lower than 0.5\%, though the UV-visible slope is better reproduced in the haze \& cloud cases, suggesting the presence of hazes is required.
The VLT and HST observations are best fitted with a haze mass flux of 10$^{-15} g.cm^{-2}.s^{-1}$.
This case is shown along with the corresponding residuals in \cref{Fig:W39best}. We highlight that these residuals are within the 3$\sigma$ of nearly all observations in the whole wavelength range covered by HST and JWST.
We therefore consider a haze mass flux of 10$^{-15} g.cm^{-2}.s^{-1}$ for WASP-39b in the rest of the study.

\section{RESULTS}
 \label{Sec:Results}

We run two sets of simulations for all planets.
The aim of the first set is to explore the effect of changing the intrinsic temperature on the thermal structure, chemical composition and haze microphysics. In the second set we explore the impact of our parameterized eddy profile keeping the high T$_{int}$ temperatures considered in the first set.

\begin{figure*}
\includegraphics[width=\textwidth]{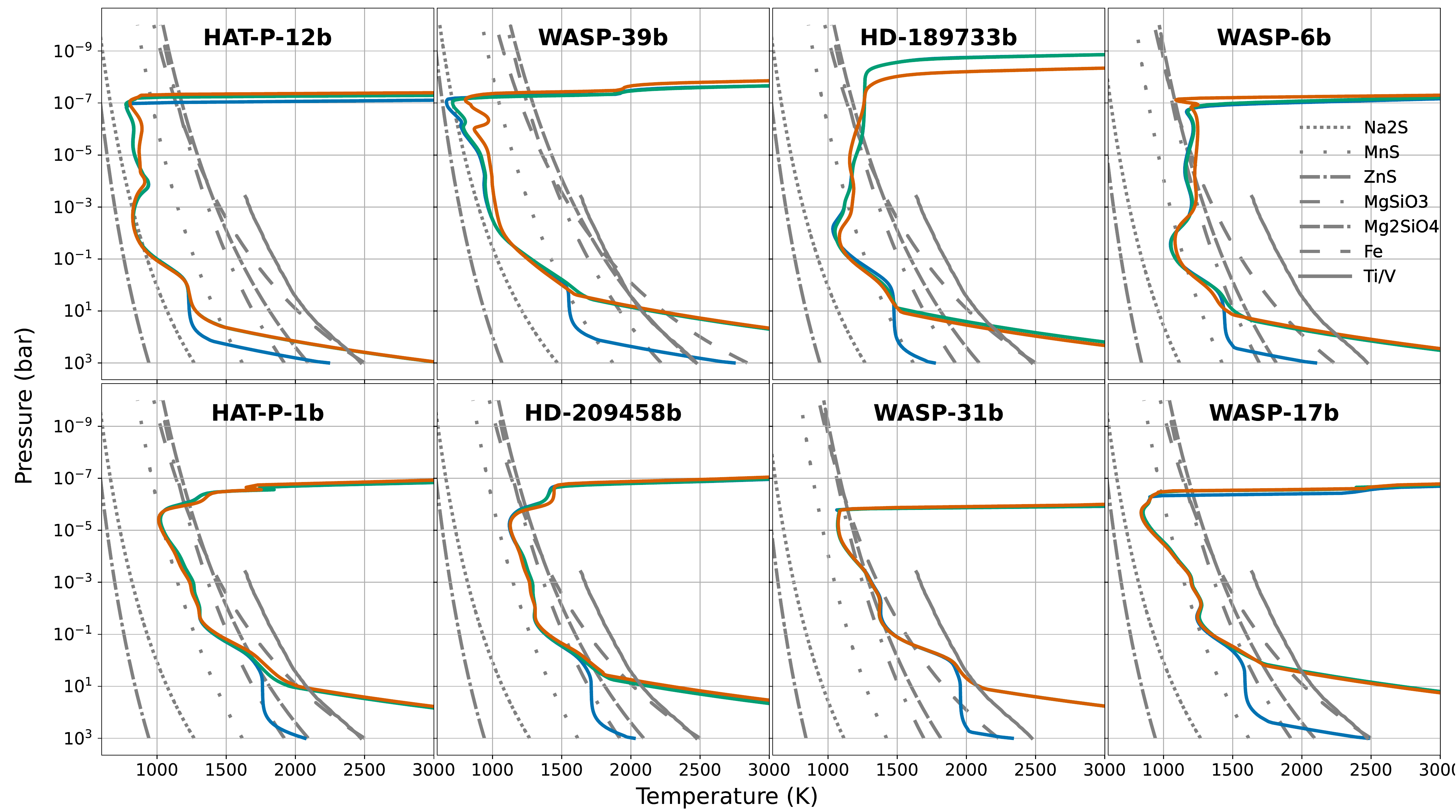}
\caption{Temperature profiles obtained with the nominal eddy in the T$_{int}$ = 100 K case (blue lines) and in the high intrinsic temperatures case (green line). Orange lines show the temperature profiles obtained with the parameterized eddy profiles and the high intrinsic temperatures. The dotted horizontal lines correspond to the location of the RCB. The grey lines are condensation curves of species expected in hot-Jupiter's atmospheres.}
\label{Fig:TempTint}
\label{Fig:TempEddy}
\end{figure*}

\subsection{Intrinsic temperature}
\label{SSec:TintRes}

\begin{figure*}
\includegraphics[width=\textwidth]{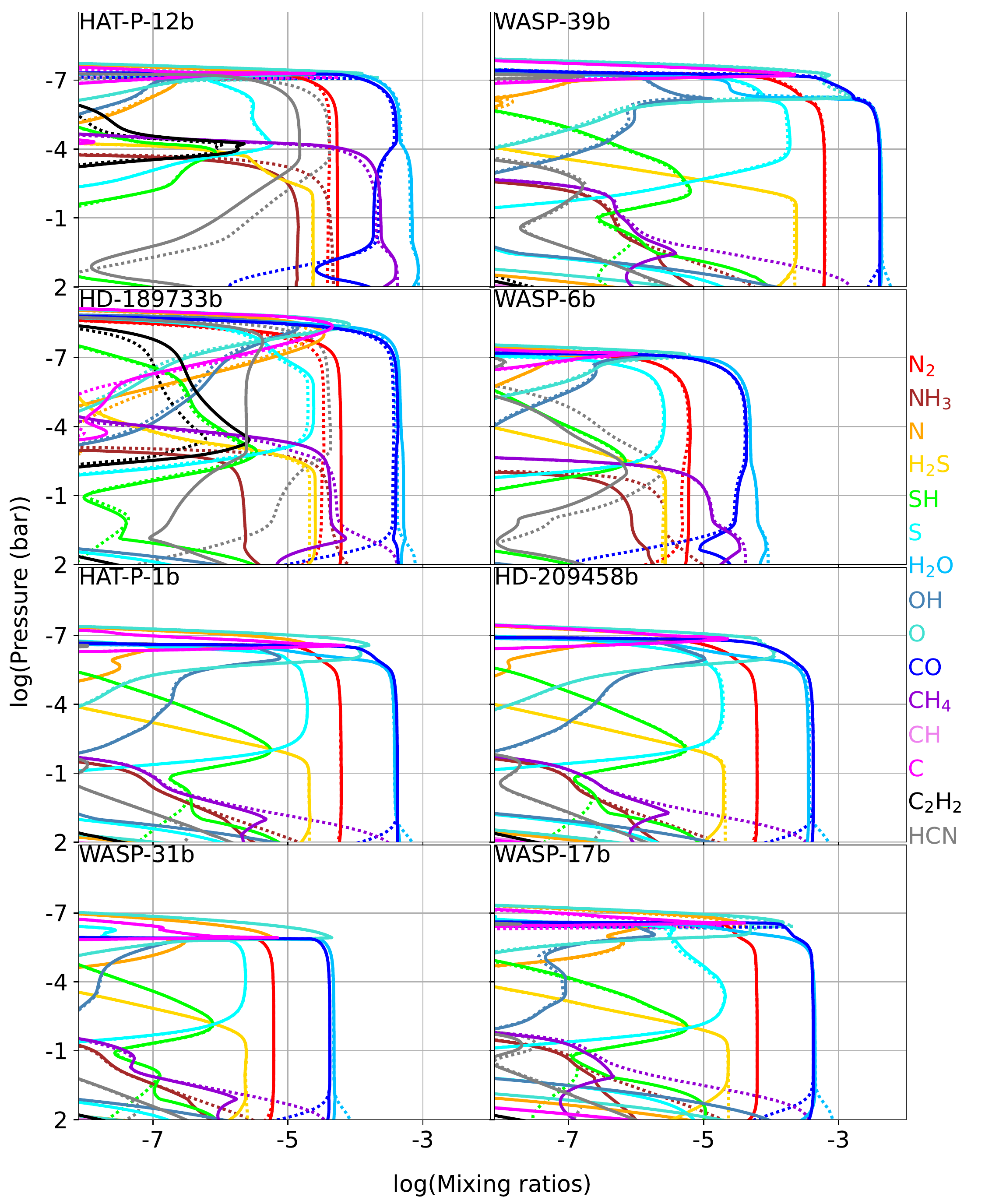}
\caption{Chemical composition profiles of the main species and their photochemical products for the T$_{int}$=100 K case (dotted lines) and for the higher intrinsic temperature case of each planet (solid lines).}
\label{Fig:ChemTint}
\end{figure*}

\begin{figure*}
\includegraphics[width=\textwidth]{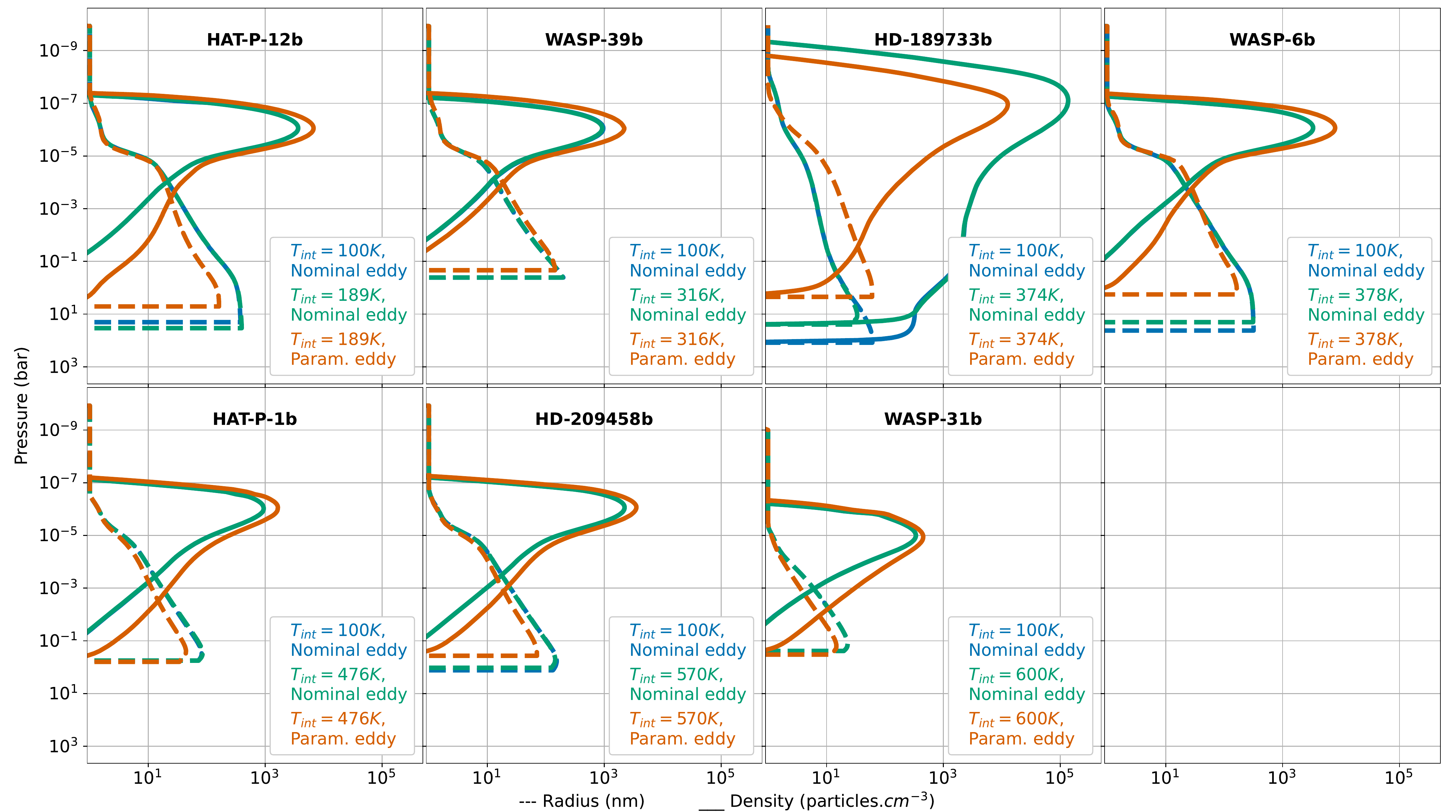}
\caption{Haze distributions using the nominal eddy profile with T$_{int}$ = 100 K (blue lines) and high intrinsic temperatures (green lines). Orange lines present the haze distribution for the parameterized eddy case with high intrinsic temperatures. Solid lines represent the particle number density, dashed lines the mean particle radius and dotted lines the haze production profile.}
\label{Fig:HazeTint}
\label{Fig:Haze}
\end{figure*}

\subsubsection{Chemistry}

A larger intrinsic temperature results in an increase of the deep atmosphere temperatures as the atmosphere follows an adiabatic lapse rate of higher entropy (\cref{Fig:TempTint}). 
The temperature at the lower boundary (1000 bar) increases by 1500 K for the coldest planet HAT-P-12b and can rise by a few thousands kelvin in the most extreme case of WASP-17b.
Moreover the higher temperatures shift the RCB to smaller pressures. The modifications of the thermal structure have further ramifications in the local chemical composition (\cref{Fig:ChemTint}). We explore here the changes brought by the higher T$_{int}$ in the deep atmosphere composition of the main gaseous species (CO, H$_2$O, N$_2$, NH$_3$, H$_2$S) and how these may affect (or not) the composition of the upper atmosphere due to quenching by the atmospheric mixing. Note that both the location of the quench levels as well as the thermochemical equilibrium abundances will change in response to the higher intrinsic temperatures. Therefore we must focus on the species mixing ratio at the quench level location in each case, referred to as quenching mixing ratio, to evaluate the implications for the upper atmosphere.

For T$_{int}$ = 100 K, the nitrogen composition is dominated by NH$_3$ in the deep atmosphere and N$_2$ at lower pressures ($<$100 bar) \citep{Lodders02}.
This qualitative picture is common among most of the studied planets, because quenching occurs at lower pressures and affects only the NH$_3$ abundance that remains in the middle atmosphere, which is always smaller than that of N$_2$ (\cref{Fig:ChemTint}).
Only for the coldest atmosphere of HAT-P-12b and the strongly mixed atmosphere of HD-189733b, the resulting deep quench levels allow ammonia to conserve a high mixing ratio and to dominate over N$_2$ to the $\sim$ 10 mbar region.
For higher intrinsic temperatures, N$_2$ starts to dominate at larger pressures.
We therefore observe a deeper N$_2$/NH$_3$ transition as we increase the intrinsic temperature, with N$_2$ dominating nitrogen chemistry down to 1000 bar.
However, we note that for the hottest planets HD-209458b, WASP-31b and WASP-17b, at pressures greater than 100 bar (not shown) the atmosphere is not dominated by NH$_3$ nor N$_2$ that are thermally dissociated, but instead we observe a domination of atomic nitrogen.
As for the  T$_{int}$ = 100 K case, quenching does not impact the global picture though it affects the ammonia upper atmosphere abundances.
For the cold planets HAT-P-12b, HD-189733b and WASP-6b, we observe lower ammonia mixing ratios compared to the T$_{int}$ = 100 K case, related to lower quenching mixing ratios.
On the other hand, for the hotter planets, the lower pressure quench levels cast the upper atmosphere immune to changes in the deep atmosphere.
We further note the case of WASP-39b that presents no major modifications of its upper atmosphere ammonia profile.
This is related to the 10$\times$solar metallicity that produces hotter deep atmosphere temperatures driving the quench levels to lower pressures, thus preserving them from the deep atmosphere temperature changes.

For sulphur chemistry, in the T$_{int}$ = 100 K case, H$_2$S is the main S-bearing species in the deep atmosphere down to 1000 bar for all planets \citep{Visscher06} and this trend is not affected by quenching.
With higher intrinsic temperatures, as the deep atmosphere temperature rises, we observe a domination of SH below the 10 bar level for HAT-P-1b, HD-209458b, WASP-17b and WASP-31b.
For the two hottest cases, WASP-17b and WASP-31b, SH is overtaken below the 100 bar level by atomic sulfur.
On the other hand, the cold planets (HAT-P-12b, HD-189733b and WASP-6b) still present a domination of H$_2$S in the deep atmosphere, except the specific case of WASP-39b due to super-solar metallicity that produces hotter deep atmosphere temperature compared to the other cold planets, thus favoring SH.
For all the studied planets, as in the T$_{int}$ = 100 K, the photolysis of H$_2$S results in the domination of atomic S in the upper atmosphere.
Apart from the deep atmosphere, the composition profiles of S-bearing species remain unaffected by the changes in intrinsic temperature owing to high quench levels.

\begin{table*}
\caption{Photochemical mass fluxes of each haze precursor ($a(b) = a\times10^{b}g.cm^{-2}.s^{-1}$) obtained in the nominal eddy and high intrinsic temperature cases. 
The fifth column contains the total mass flux summing all the individual contributions, the sixth contains the value in the T$_{int}$ = 100 K case and the seventh contains the ratio of the two.
The penultimate column shows the inferred mass flux from the observations and the last presents the corresponding yield.
}
\begin{adjustbox}{max width=\textwidth}
\begin{tabular}{c|cccc|c|c|cc|cH}
\hline
Planets			&	$CH_4$	&	$CO$	&	$HCN$	&	$C_2H_2$&	High T$_{int}$	&	T$_{int}$ = 100 K	& 	Ratio		&	 Mass flux  	& 	Formation yield (\%)		& Metallicity	\\
\hline
HAT-P-12b		&	1.5(-19)	&	5.0(-14)	&	9.1(-12)	&	5.2(-14)	&	9.2(-12)		&	2.4(-11)			&	0.4		&	1.0(-14)		&		0.11			&		1	\\
WASP-39b		&	1.4(-28)	&	1.7(-11)	&	4.6(-14)	&	1.2(-19)	&	1.7(-11)		&	1.7(-11)			&	1.0		&	1.0(-15) 		&		0.006		&		1	\\
HD-189733b		&	1.6(-19)	&	1.6(-13)	&	1.2(-11)	&	1.6(-11)	&	2.8(-11)		&	1.4(-10)			&	0.2		&	9.0(-12)		&	 	32			&		1	\\
WASP-6b			&	3.3(-20)	&	7.7(-16)	&	6.5(-14)	&	1.3(-18)	&	6.6(-14)		&	2.6(-13)			&	0.3		&	1.0(-14)		&		15			&		0.1	\\
HAT-P-1b			&	6.7(-22)	&	2.6(-12)	&	2.5(-13)	&	2.4(-17)	&	2.9(-12)		&	2.9(-12)			&	1.0		&	7.0(-16)		&		0.024 		&		1	\\
HD-209458b		&	2.4(-23)	&	1.4(-13)	&	2.5(-15)	&	2.5(-18)	&	1.4(-13)		&	1.6(-13)			&	0.9		&	3.3(-15)		&		2.1			&		1	\\
WASP-31b		&	2.1(-20)	&	1.4(-13)	&	2.8(-14)	&	6.6(-19)	&	1.7(-13)		&	1.9(-13)			&	0.9		&	1.8(-17)		&		0.01			&		0.1	\\
WASP-17b		&	1.7(-21)	&	1.6(-12)	&	4.8(-13)	&	1.7(-16)	&	2.1(-12)		&	5.1(-13)			&	4.1		&	$<$1.0(-16)	&		$<$0.002		&		1	\\
\hline
\end{tabular}
\end{adjustbox}
\label{Tab:PhotoFluxTint}
\end{table*}

Carbon and oxygen chemistry are strongly connected since they share a common dominating species: carbon monoxide.
For T$_{int}$ = 100 K, we observe a domination of CH$_4$ (for carbon chemistry) and H$_2$O (for oxygen chemistry) at large pressures, and a domination of CO (for both carbon and oxygen chemistry) at lower pressures (\cref{Fig:ChemTint}).
However, CO mixing ratio is constrained by the elemental abundance of carbon based on the solar C/O ratio that we assume \citep[{[C/O]} = 0.457,][]{Lodders10}.
We therefore observe similar CO and H$_2$O mixing ratios above the H$_2$O/CO transition, since the whole carbon content is consumed when approximately half of the oxygen content is used, so that CO cannot contain more than half of the total oxygen elemental abundance, the other half stored in water.
We observe methane dominating below 100 bar for the four hottest planets (HAT-P-1b, HD-209458b, WASP-31b and WASP-17b), up to the 10 bar level for HD-189733b and WASP-6b and up to 1 bar for HAT-P-12b.
We note the exception of WASP-39b with a super-solar metallicity that drifts the CO - CH$_4$ iso-composition curve to larger pressures \citep{Fortney20}, thus allowing a domination of CO down to 1000 bar.
For HAT-P-12b, we further note that the deep quench levels allow CH$_4$ to remain as abundant as CO up to the 10 mbar level.
For the other planets, the quench levels are higher than the CH$_4$/CO transition.
On the other hand, we note the impact of photolysis of CO and H$_2$O in the upper atmosphere that leads a domination of atomic carbon and oxygen above the 1 µbar altitude.

As the intrinsic temperature increases, the CH$_4$/CO transition moves to larger pressures leading to a domination of carbon monoxide down to 1000 bar for most planets.
As a result, the water abundance in the deep atmosphere decreases, compared to the T$_{int}$ = 100 K case, down to values similar to carbon monoxide.
Therefore, CO and H$_2$O abundances are about the same in the whole range of the atmosphere simulated and remain unaffected by transport.
For HAT-P-12b, the picture is slightly more complex due to the relatively cold equilibrium temperature of this planet.
Indeed, \cite{Fortney20} founds that, for rather cold planets, CO dominates at high pressures and methane at lower pressures, while for hotter planets, the behavior is reversed.
HAT-P-12b, with T$_{eq}$ = 956 K, is close to the transition region between these two behaviors, therefore, though CO actually dominates the deep atmosphere, we observe the region between 1 and 100 bar where methane locally takes over carbon monoxide.
Despite its slightly hotter temperatures, WASP-6b presents a similar behavior, related to the sub-solar (0.1$\times$) metallicity used for this planet.
Indeed, a sub-solar metallicity drifts the CO - CH$_4$ iso-composition curve to higher temperatures allowing the formation of a local CH$_4$ dominated region \citep{Fortney20}.
As a consequence, these two planets also present a domination of water over CO in the deep atmosphere while the other planets have similar CO and H$_2$O abundances.
Finally, for the two hottest planets WASP-17b and WASP-31b, the region below 100 bar is partly dominated by O and OH for oxygen chemistry and C and CH for carbon chemistry owing to the water and methane thermal dissociation.
As with the sulfur chemistry, the O-bearing species, as well as, CO and CH$_4$ upper atmosphere profiles remain unaffected by the changes in intrinsic temperature.

As an overall picture, we observe that for planets with equilibrium temperatures larger than $\sim$1300 K, the upper atmosphere remains unaffected by the deep atmosphere changes brought by increasing the intrinsic temperature, owing to high up quench levels.
On the other hand, for the cold planets HAT-P-12b, WASP-6b and WASP-39b, the low temperatures allow for quench levels deeper down in the atmosphere, which can therefore be impacted by the larger deep atmosphere temperatures.
While CH$_4$ and CO conserve sufficiently high quench levels preserving the upper atmosphere profiles of these species from the deep atmosphere changes, the other haze precursors present deeper quench levels, therefore their abundances can be affected.
As previously discussed, when increasing the intrinsic temperature, the NH$_3$/N$_2$ balance favors N$_2$, and ammonia quenching mixing ratio becomes $\sim$3$\times$ smaller than in the 100 K case.
As the HCN upper atmosphere abundance is related to its production around 1 mbar from the photodissociation of NH$_3$ and CH$_4$, we observe weaker HCN abundances in HAT-P-12b and WASP-6b atmospheres (\cref{Fig:ChemTint}).
Therefore, since HCN is the main haze precursor for planet equilibrium temperatures below 1300 K \citep{Arfaux22}, we observe a weaker haze precursors photolysis mass flux for these planets (\cref{Tab:PhotoFluxTint}).
For WASP-39b however, the 10$\times$solar metallicity used allows CO to be the dominating haze precursor.
Since this species is not strongly affected by changes in intrinsic temperature, the haze precursors photolysis mass flux remains roughly the same between the two T$_{int}$ cases for WASP-39b.

In the particular case of HD-189733b, not only the cold temperatures but also the strong nominal eddy profile used (100$\times$ larger than for the other planets) produce quench levels deeper down in the atmosphere.
As a result, we obtain for HD-189733b a NH$_3$ quenching mixing ratio 40$\times$ smaller in the T$_{int}$ = 374 K case than in the 100 K case, resulting in a similar drop in HCN upper atmosphere mixing ratio (\cref{Fig:ChemTint}).
As the CH$_4$ quenching mixing ratio is preserved in changing the intrinsic temperature, its upper atmosphere profile remains unchanged and its photolysis to form CH$_3$ stays as powerful.
Since a smaller part of the produced CH$_3$ reacts to form HCN (as there is less nitrogen to react with), CH$_3$ will be mainly lost to form C$_2$H$_4$ and C$_2$H$_6$, which will dissociate in C$_2$H$_2$.
We therefore observe larger C$_2$H$_2$ upper atmosphere abundances, further increasing its contribution to the haze precursors photolysis.
Thus, while the C$_2$H$_2$ contribution to haze formation is negligible for all other planets and for the 100 K case for HD-189733b, C$_2$H$_2$ becomes the main haze precursors in the 374 K case for HD-189733b.
However the total mass flux from the photolysis of the haze precursors remains lower in T$_{int}$ = 374 K case than in the 100 K case (\cref{Tab:PhotoFluxTint}) due to the weaker HCN contribution.

\subsubsection{Haze distribution and transit spectrum}

\begin{figure*}
\includegraphics[width=\textwidth]{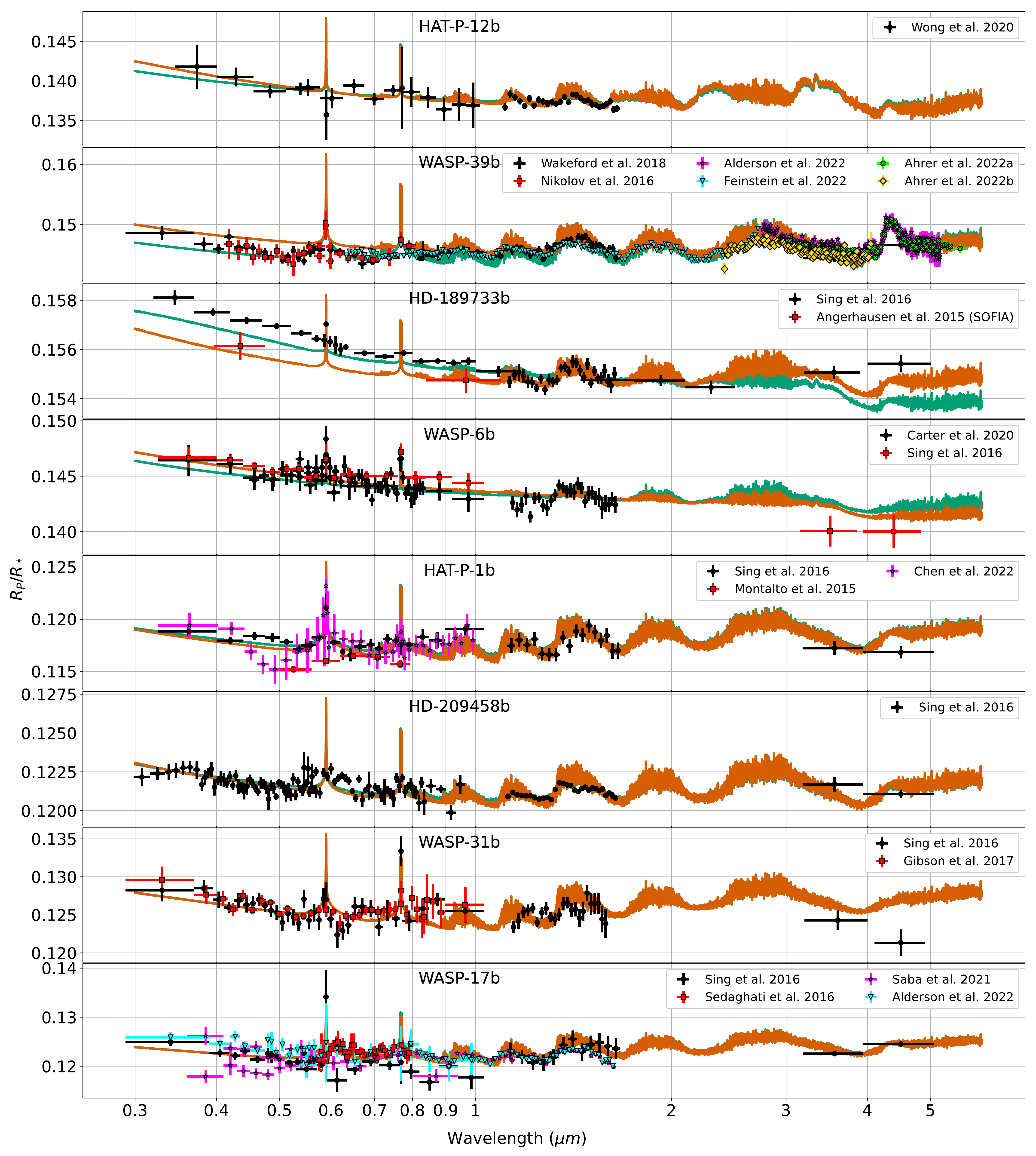}
\captionof{figure}{Spectra obtained for the high T$_{int}$ case with the nominal eddy profile (green lines) and the parametrized eddy profiles (orange lines).
The crosses are observations, the black are taken with the Hubble Space Telescope and Spitzer.
The others are additional space- or ground-based observations.
A detailed list of the observations presented here is available in Tab. 2 of \protect\cite{Arfaux22}.
We additionally present the recent observations by the JWST for WASP-39b \protect\citep{JWST22,Alderson22b,Ahrer22,Feinstein22,Rustamkulov22}.
}
\label{Fig:SpectraEddy}
\end{figure*}

The upper atmosphere temperature profiles are not affected by the change of intrinsic temperature, thus resulting in weak changes of the haze distributions (\cref{Fig:HazeTint}).
The only notable modification is that particle sublimation starts earlier during particle settling (i.e. at lower pressures). This does not have any impact on the thermal structure.
Since our calculations demonstrate small modifications in the temperature, main radiatively-active species abundances and haze distribution in the region of the atmosphere probed by the observations, our simulated transit spectra for the case of high T$_{int}$ (\cref{Fig:SpectraEddy}) demonstrate negligible changes from those derived in our previous study and used for the evaluation of the haze mass flux in each planet case \citep{Arfaux22}. 
As a consequence, we can still consider the assumed haze mass fluxes as valid.
On the other hand, for the cold planets (with equilibrium temperatures lower than 1300 K), the decrease in haze precursors photolysis mass flux due to lower HCN abundances indicates that the haze formation yields for these planets could be larger than those inferred in our previous work \citep{Arfaux22}.
We obtain differences in the yields ranging from a factor of 3 for HAT-P-12b to a factor of 5 for HD-189733b (\cref{Tab:PhotoFluxTint}).
Therefore, fixing the yield and deriving the haze mass flux from the precursors photolysis mass flux could have an impact on the resulting particle size distribution.
Thus, in the next section we explore further this possibility and how the changes in the haze precursors abundances could affect the resulting haze properties.

\subsection{Eddy profile}
\label{SSec:EddyRes}

In \cref{SSec:TintRes}, we demonstrated that changing the intrinsic temperature can affect the upper atmosphere haze precursors for planets with equilibrium temperatures lower than 1300 K. This has direct implications on the haze formation, thus on the resulting microphysical properties of the particles. In order to explore these implications we will use the haze formation yield instead of the haze mass flux in the rest of our analysis, where we test our eddy profile parameterization. This way we can explore both the direct impact of the parameterization on the haze distribution as well as its implications on the precursors distribution that controls the haze production through their photolysis. 
This approach does not optimize the fits to the observations but allows to explore the feedbacks between the chemistry and the microphysics, and their ramifications on the transit spectra.
Therefore, in the second step of this work, we test our parameterized eddy with our self-consistent model keeping the high intrinsic temperatures explored in the previous section and fixing the haze formation yield to the value retrieved for each planet (\cref{Tab:PhotoFluxTint}).

\subsubsection{General results}

\begin{figure*}
	\centering
		\includegraphics[angle=90,height=0.9\textheight]{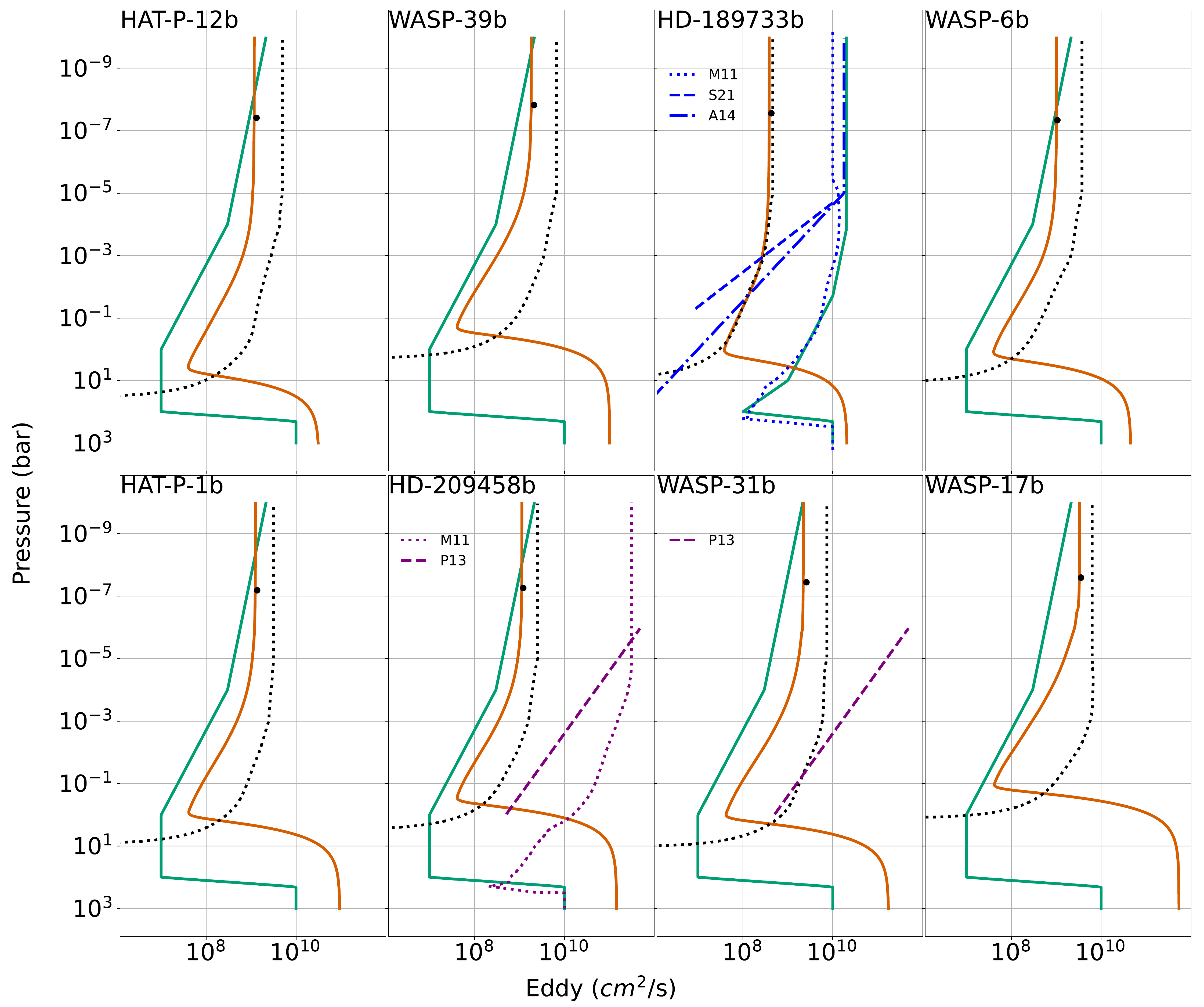}
	\caption{Comparison between the nominal (\protect\cite{Moses11} for HD-189733b) eddy profiles (green lines) and the parametrized eddies (orange lines).
	The dotted black lines are profiles calculated with \protect\cite{Zhang18a} eddy diffusion parameterization and using the \protect\cite{Komacek19} vertical wind parameterization.
	HD-189733b, HD-209458b and WASP-31b panels present additional profiles from GCM results (M11: \protect\cite{Moses11},  P13: \protect\cite{Parmentier13},  A14: \protect\cite{Agundez14}, S21: \protect\cite{Steinrueck21}).
	The black dots correspond to the homopause.}
	\label{Fig:SelfEddies}
	\label{fig:eddies}
\end{figure*}

\begin{figure*}
	\centering
		\includegraphics[width=\textwidth]{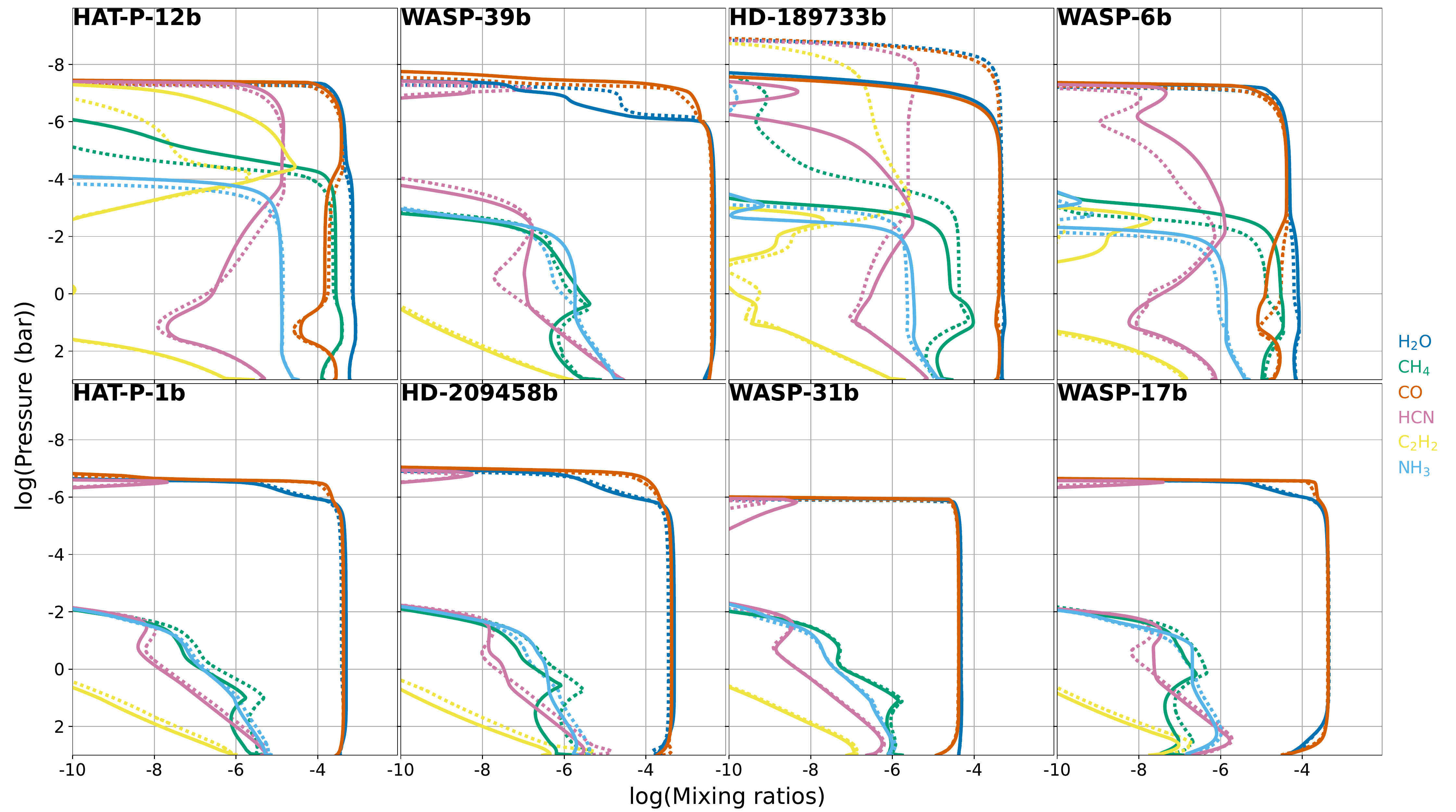}
	\caption{Chemical composition profiles of the haze precursors, as well as, NH$_3$ and H$_2$O, with the nominal eddy profile (dotted lines) and the parameterized eddy profiles (solid lines).}
	\label{Fig:ChemEddy}
\end{figure*}

In most cases, our parameterized eddy profiles are similar in shape and magnitude to the nominal profiles previously used in our calculations (\cref{Fig:SelfEddies}), with variations lower than one order of magnitude above the RCB.
We however note that our parameterized eddies are systematically higher than the nominal profiles assumed previously.
At the bottom boundary, owing to the large temperatures produced by the high T$_{int}$, the parameterized eddies are larger than the nominal, with variations ranging from one order of magnitude for the coldest planet HAT-P-12b, up to two orders of magnitude for the hot planet WASP-17b. Moreover, the high intrinsic temperatures shift the RCB to lower pressures, thus resulting in expanded convective regions.
This keeps convective mixing efficient relatively high up in the atmosphere above the 10 bar altitude and up to the 0.1 bar level in the most high T$_{int}$ cases, while in the nominal profiles the convective region was limited to pressures larger than $\sim$100 bar.
Therefore, in the region from 100 bar to the RCB, the parameterized eddy presents values larger than the nominal by 3 orders of magnitude and up to 5 orders of magnitude for WASP-17b.
Finally, we note the particular case of HD-189733b for which our parameterized eddy profile is two orders of magnitude lower above the RCB compared to the nominally used for this planet, though it remains larger below the RCB as for the other planets.

For comparison, we show in dotted black line in \cref{Fig:SelfEddies} the profiles derived with \cite{Zhang18a} parameterization of the eddy diffusion and assuming the vertical wind speed parameterization from \cite{Komacek19}, using the RCB as the lowest pressure where the dayside-nightside temperature difference is zero. We use a chemical timescale of 1.6$\times$10$^5$s as in \cite{Komacek19}. The comparison to the profiles provided with our parameterization indicates a rather good agreement with differences within an order of magnitude above the 1 mbar altitude in most cases. At higher pressures, however, \cite{Zhang18a}/\cite{Komacek19} and our parameterizations diverge owing to the lack of convective regime in the former.

The larger eddy diffusion provided by our parameterization in the deep atmosphere enhances the transport of chemical species drifting the quench levels to higher pressures.
This modification can have various implications depending on the thermochemical equilibrium profiles of the species in the different studied atmospheres.
The chemical response to the larger eddy can therefore change from a planet to another and the details will be explored below for each planet individually.
Despite these individual variations, a global picture appears for the haze precursors with CO and CH$_4$ relatively unaffected and HCN and C$_2$H$_2$ demonstrating higher abundances, related to a more efficient redistribution of these species from their source region under the effect of the stronger eddy diffusion (\cref{Fig:ChemEddy}). As a result, while CO and CH$_4$ contributions to the haze precursors remain almost constant, the increase of HCN and C$_2$H$_2$ abundances enhances the haze production (\cref{Tab:PhotoFluxEddy}).
This picture reverses for the particular case of HD-189733b for which our parameterization provides a weaker eddy compared to the nominally used \cite{Moses11} profile, and results in a weaker haze formation.

\begin{table*}
\caption{
Photochemical mass fluxes of each haze precursor ($g.cm^{-2}.s^{-1}$) obtained in the parameterized eddy and high intrinsic temperature case, integrated above 10$^{-5}$ bar. 
The fifth column contains the total mass flux summing all the individual contributions and the sixth contains the ratio of the total mass flux to that retrieved in the nominal eddy and high T$_{int}$ case (\cref{Tab:PhotoFluxTint}).
}
\begin{adjustbox}{max width=\textwidth}
\begin{tabular}{c|cccc|c|HcHHH}
\hline
Planets			&	$CH_4$	&	$CO$	&	$HCN$	&	$C_2H_2$&	Total		&	Previous (nominal)	& 	Ratio		&	Retrieved  	& 		Yield (\%)		& Metallicity	\\
\hline
HAT-P-12b		&	8.5(-17)	&	5.0(-14)	&	1.2(-11)	&	2.2(-11)	&	3.4(-11)	&	9.2(-12)			&	3.7		&	1.0(-14)		&		0.03			&		1	\\
WASP-39b		&	8.2(-30)	&	8.2(-11)	&	5.6(-15)	&	4.1(-20)	&	8.2(-11)	&	1.7(-11)			&	4.8		&	1.0(-15)		&		$<$0.2		&		1	\\
HD-189733b		&	4.2(-21)	&	1.9(-16)	&	2.4(-13)	&	4.7(-23)	&	2.4(-13)	&	2.8(-11)			&	0.009	&	5.0(-12)		&	 	6.4			&		1	\\
WASP-6b			&	2.4(-19)	&	1.4(-15)	&	5.0(-13)	&	2.6(-18)	&	5.0(-13)	&	6.6(-14)			&	7.6		&	1.0(-14)		&		15			&		0.1	\\
HAT-P-1b			&	1.8(-22)	&	8.8(-12)	&	4.7(-13)	&	1.0(-17)	&	9.3(-12)	&	2.9(-12)			&	3.2		&	7.5(-16)		&		0.03 			&		1	\\
HD-209458b		&	7.9(-24)	&	4.7(-13)	&	5.8(-15)	&	1.5(-18)	&	4.8(-13)	&	1.4(-13)			&	3.5		&	1.0(-14)		&		2.2			&		1	\\
WASP-31b		&	1.1(-18)	&	9.3(-13)	&	3.4(-13)	&	1.0(-16)	&	1.4(-12)	&	1.7(-13)			&	7.6		&	1.0(-12)		&		0.01			&		0.1	\\
WASP-17b		&	2.2(-22)	&	3.5(-12)	&	1.1(-12)	&	3.6(-17)	&	4.6(-12)	&	2.1(-12)			&	2.2		&	$<$1.0(-16)	&		$<$0.003		&		1	\\
\hline
\end{tabular}
\end{adjustbox}
\label{Tab:PhotoFluxEddy}
\end{table*}

We observe two different behaviors among the studied planets based on the haziness conditions.
The transit spectrum of planets presenting an apparent absence of haze in their atmospheres is not strongly impacted and presents negligible modifications.
Indeed, the major chemical species affecting the transit (H$_2$O, CO, Na, K) are well mixed in the probe pressure range, therefore, changing the strength of diffusion for these atmospheres has negligible influence.

For hazy planets, changing the eddy profile has a strong influence on the haze distribution that has further ramifications on the thermal structure of the planet and the transit spectrum.
A strong vertical mixing reduces the particle transport timescale, thus hampering their coagulation. A large eddy diffusivity is therefore expected to result in small and numerous particles while a weak eddy would produce larger and fewer particles.
This behavior is observed in \cref{Fig:Haze} with larger densities and smaller mean particle radii with our slightly larger parametrized eddy diffusion profile relative to the nominal. In addition, the enhanced haze production (\cref{Tab:PhotoFluxEddy}) further raises the particle number density, especially impacting the upper atmosphere where the mean particle size remains small and close to the nominal eddy case.
The radiatively active species are not strongly impacted by this change of eddy diffusion profile, resulting in weak modifications of their impact on the temperature.
However, for the most hazy planets (WASP-6b and HAT-P-12b), the more numerous particles result in larger local temperatures.
In the specific case of HD-189733b, the trend on the thermal structure is reversed owing to the fewer particles relative to the nominal eddy case.

For all planets, we note modifications in the thermosphere boundary altitude (\cref{Fig:TempEddy}) related to the slightly larger parameterized eddy diffusivity in the upper atmosphere (reversed for HD-189733b).
However, this modification is negligible in most cases and does not impact the transit spectra.

\subsubsection{HAT-P-12b}

The slightly larger eddy mixing provided by our parameterization enhances CH$_4$ abundances in the middle atmosphere (between 1000 and 0.1 bar) but leaves NH$_3$ nearly unchanged (\cref{Fig:ChemEddy}).
As a result, the C$_2$H$_2$ production at 0.1 mbar is enhanced since more material is available from CH$_4$ photochemistry.
C$_2$H$_2$ is then transported to the haze production region via eddy mixing where it participates to the formation of haze particles and becomes the main haze precursor for HAT-P-12b leading to a $\sim$4 times larger haze precursors photolysis mass flux. The larger haze mass flux results in more numerous particles  (\cref{Fig:Haze}). 
However, the impact of the stronger parameterized eddy partly counterbalances this effect of higher haze mass flux, resulting in only twice more particles in the production region and a mean particle radius close to the nominal eddy value down to $\sim$ 0.1 mbar. The more numerous particles and similar particle size increases the optical depth in the upper atmosphere. The higher haze optical depth results in a local increase of the atmospheric temperature by $\sim$60 K (\cref{Fig:TempEddy}). The increase in optical depth in the production region and the larger atmospheric scale height due to the higher local temperature lead to a steeper slope of the UV-visible transit spectrum (\cref{Fig:SpectraEddy}). 
Below the $\sim$ 0.1 mbar altitude, the larger eddy hampers the particle coagulation yielding two times smaller particles at 10 mbar altitude, further enhancing the particle number density compared to the nominal eddy case. However, this region is not probed by the transit observations, therefore the changes in particle distribution are not impacting the transit spectrum.

\subsubsection{WASP-39b}

WASP-39b presents an increase of NH$_3$ mixing ratio by a factor of $\sim$4 relative the the nominal eddy case in the region from 1 to 0.01 bar, related to the deeper quenching of this species.
However, the material produced by the photodissociation of ammonia near 0.01 bar is lost to N$_2$ and does not lead to any enhancement of the production of HCN.
On the other hand, the larger CO abundance near 0.1 µbar, where most of CO photolysis happens owing to the weak penetration depth of the required high-energy photons, results in a factor of $\sim$9 increase of the effective haze mass flux.
The larger haze mass flux results in much more numerous particles thus increasing their collision probability, and then enhancing the particle coagulation rates.
This stronger particle coagulation outbalances the hampering of the coagulation related to the powerful eddy.
As a result, we observe both larger and more numerous particles in the parameterized eddy case resulting in much larger haze optical depths relative to the nominal eddy case.
The enhanced haze opacity produces temperatures $\sim$50 K hotter between 10 and 0.01 mbar and up to $\sim$200 K hotter in the haze production region, thus increasing the atmospheric scale height.
This enhanced atmospheric expansion as well as the larger haze optical depth result in larger transit depths in the UV-visible region of the transit spectrum (\cref{Fig:SpectraEddy}), thus suggesting an even smaller haze formation yield.

\subsubsection{HD-189733b}

HD-189733b presents the largest differences between the parameterized and nominal eddy profiles.
Indeed, the nominal profile was not down-scaled for this planet, resulting in eddy mixing profile two orders of magnitude larger than for the other planets.
These large eddy diffusion coefficients are however not reproduced with our common parameterization, which provides values similar to those obtain for the other planets.
In \cref{Fig:SelfEddies}, we present additional parameterizations for HD-189733b based on GCMs \citep{Agundez14,Steinrueck21}, in addition to the nominal profile \citep{Moses11}.
\cite{Agundez14} and \cite{Steinrueck21} use the tracer distribution method, while \cite{Moses11} assumes the rms vertical velocity parameterization based on GCM results from \cite{Showman09}.
The estimations based on tracers assume a pressure power law ($K = \alpha P^{-\gamma}$) similar to our assumption.
The main difference is on the $\gamma$ coefficient: the values retrieved by \cite{Agundez14} and \cite{Steinrueck21} are 0.65 and 0.9, respectively,
compared to our value of 0.5 based on gravity waves break down \citep{Chamberlain78}.
Nevertheless, these calculations do not account for the effect of gravity waves, but for large scale motions within atmospheric cells, which may explain this disagreement.
Nevertheless, the order of magnitude provided by these parameterizations is in agreement with our parameterization in the middle atmosphere.

Our parameterized eddy profile follows the same shape as the nominal profile but with values a hundred times smaller above the RCB level, located near 1 bar.
In the deep atmosphere, the parameterized eddy is larger than the nominal profile, due to the high T$_{int}$ assumed.
The two profiles cross each other around 10 bar, which corresponds to the quench level of most species.
We therefore have small modifications of the quenching mixing ratio of the haze precursors.
However, the weaker eddy in the upper atmosphere leads to lower CH$_4$ mixing ratios and then smaller photolysis rates (\cref{Fig:ChemEddy}).
Therefore, in addition to the weaker contribution of methane to the haze precursors, the resulting smaller production of CH$_3$ affects the C$_2$H$_2$ abundance and, to a lesser extent, the HCN local formation around 1 mbar.
In addition, the weaker transport of these species from their production region to the rest of the atmosphere results in lower overall abundances.
Furthermore, the weak eddy provides a lower homopause altitude, further reducing the haze precursors column density in the haze production region above 10 µbar.

The lower parameterized eddy therefore results in a much smaller haze formation with a haze mass flux of 6.4x10$^{-14} g.cm^{-2}.s^{-1}$ against 9x10$^{-12} g.cm^{-2}.s^{-1}$ in the nominal eddy case, both corresponding to the same formation yield of $\sim$30 \%.
As a result of this much smaller haze production, the particle number density drops by a factor of ten at the production altitude and a factor of a hundred at 0.1 bar compared to the nominal eddy case (\cref{Fig:Haze}).
The much smaller number density results in a less efficient coagulation and then smaller particles.
Although this latter effect is partly counterbalanced by the much weaker eddy that effectively enhances the particle coagulation, we observe smaller particles in the haze formation region around 0.1 µbar.
Below 1 µbar, the coagulation starts to have a strong impact and quickly produces large particles reaching radii of 20 nm at 1 mbar and up to 60 nm at 1 bar.

The reduced haze mass flux, results in a decrease of the haze opacity in the whole atmosphere, providing a weaker UV-visible transit depth in the spectrum (\cref{Fig:SpectraEddy}).
We also note that this smaller haze opacity results in a less muted water band, as well as, modifications of the thermal structure (\cref{Fig:TempEddy}).
Indeed, in addition to the lower thermosphere related to the weaker eddy, we note slightly cooler temperatures ($\sim$50 K) in the haze production region, owing to the weaker haze opacity, and slightly hotter ($\sim$50 K) between 0.1 and 100 mbar, as significantly more radiation can reach these pressures compared to the nominal eddy case.

The resulting transit spectrum is significantly under predicting the HST/STIS observations, which however could be largely affected by stellar effects \citep{McCullough14,Arfaux22}. On the contrary, the new transit spectrum is consistent with the SOFIA observations \citep{Angerhausen15}, while it results in a much improved agreement with the HST/WFC observations of the water band (\cref{Fig:SpectraEddy}). These characteristics suggest that the parameterized eddy could be an improvement over the nominal case used before, but we need better constraints to resolve this issue of stellar contamination in the HD189733b observations. Notwithstanding these limitations, we discuss further below the necessary modifications in our parameterized eddy profile to fit the STIS observations.

\subsubsection{WASP-6b}

As for HAT-P-12b, our parameterized eddy profile for WASP-6b is rather close to the nominally used profile above the RCB, though slightly larger within a factor of three (\cref{Fig:SelfEddies}).
However, below the 1 bar level our parameterized K$_{ZZ}$  provides a significantly stronger mixing compared to the nominal profile, thus enhancing the quenching of ammonia and methane and resulting in about twice larger mixing ratio for these species between 10 and 0.001 bar.
We therefore observe a stronger production of HCN and C$_2$H$_2$ near 10 mbar, with a factor of two increase for HCN and a factor of 100 for C$_2$H$_2$ (\cref{Fig:ChemEddy}).
The photolysis of NH$_3$ is controlling the abundance of HCN, therefore the twice larger NH$_3$ abundance results in twice larger HCN abundance at its production level.
On the contrary, methane photolysis results in multiple species through a complex network of reactions and eventually yields as small C$_2$H$_2$ production.
Therefore, a factor of only 2 increase in methane abundances has a non-linear impact on the C$_2$H$_2$ density and results in a much stronger production of this molecule.
On the other hand, acetylene is strongly photo-dissociated below the haze production region where the haze mass flux is evaluated (above the 10 µbar level) and therefore, this strong enhancement of C$_2$H$_2$ abundance has a negligible impact on the haze production.
The more efficient transport and the twice larger HCN mixing ratio at its production level result in an important increase of this species abundance in the upper atmosphere (above its production region) with a mixing ratio up to one order of magnitude larger compared to the nominal eddy case.
HCN remains therefore the main haze precursor and its contribution to the haze formation results in an almost eight times larger haze mass flux compared to the nominal eddy case (\cref{Tab:PhotoFluxEddy}).

This strong haze enhancement observed for WASP-6b leads to an increase of the mean particle radius (\cref{Fig:Haze}) by 12\% above the 1 µbar level and by 40\% in the region from 1 to 0.01 mbar, relative to the nominal eddy case.
The particle size increase in the 1 to 0.01 mbar region only appears in WASP-6b and is related to the larger haze production enhancement compared to that evaluated for the other planets.
Indeed, for most other hazy planets, the particle coagulation enhancement due to the larger haze mass flux is outbalanced by the stronger eddy mixing.
However, the much stronger haze enhancement observed in WASP-6b results in an more efficient coagulation of the particles in the production region relative to the nominal eddy case.
These larger and more numerous particles increase the haze opacity, leading to upper atmosphere temperatures hotter by $\sim$50 K (\cref{Fig:TempEddy}), thus resulting in a larger scale height.
The larger opacities in the probed region and the atmospheric expansion related to a hotter upper atmosphere, provide an increase of the transit depth in the UV-visible transit spectrum, as well as, a steeper slope in this region of the spectrum (\cref{Fig:SpectraEddy}).
As a consequence, the fit to the HST-STIS observations is improved, however, the enhanced muting of the water band, related to stronger haze opacities, results in a worse fit of the HST-WFC3 observations.

\subsubsection{HAT-P-1b}

Due to its rather high atmospheric temperatures, HAT-P-1b presents small quenching mixing ratios for most haze precursors (HCN, CH$_4$ and C$_2$H$_2$), resulting in weak upper atmosphere abundances for these species under the nominal eddy profile.
However, the dissociation of H$_2$O, CO and N$_2$ at the base of the thermosphere (around 1 µbar) leads a local production of HCN (\cref{Fig:ChemEddy}), which represents the main source of HCN contribution to the haze formation in our calculations.
For CO the high-energy photons required for photodissociation are quickly extinct, also resulting in a contribution of CO to the haze formation strongly localized above 1 µbar.
The larger eddy profile provided by our parameterization for HAT-P-1b enhances the CO and N$_2$ mixing ratios in that pressure range, thus resulting in a similar enhancement of HCN.
We therefore observe a twice larger photolysis mass flux by HCN and three times larger by CO (\cref{Tab:PhotoFluxEddy}).  
This enhanced contribution of HCN and CO produces a three times larger haze mass flux resulting in  an almost twice larger particle density (owing to the counterbalancing of the stronger eddy profile) in the haze production region (\cref{Fig:Haze}).
Below the production region, the larger eddy mixing hampers the particle coagulation leading to twice smaller particles at 0.1 bar, further increasing the particle density up to a factor of 4, relative to the nominal eddy case.
Despite the strong variation in HAT-P-1b haze distribution, we see a rather weak modification of its transit spectrum between the two eddy cases.
This is related to the small amount of haze barely sufficient to have a significant impact on the spectrum, though we note a slightly smaller transit depths in the visible with the parameterized eddy.
Indeed, the low haze mass flux allows to probe large pressures where particles are large and few.
In this region, the increase of particle number density has less impact on the haze opacity than the variation of the particle size.
We therefore observe lower haze opacities in the region of atmosphere probed by the STIS observations of HAT-P-1b related to the smaller particles produced by the stronger eddy profile.

\subsubsection{HD-209458b}

In \cref{Fig:SelfEddies}, we show additional eddy profiles derived for HD-209458b by \cite{Moses11} based on the rms vertical velocity, and by \cite{Parmentier13} based on tracer distribution.
Our parameterized eddy profile for HD-209458b is rather close to the nominal above the RCB (\cref{Fig:SelfEddies}). In comparison the  \cite{Moses11} eddy profile, based on the rms vertical velocity, is $\sim$ two orders of magnitude larger, while the \cite{Parmentier13} eddy profile follows a $\gamma$ = 0.5 behavior at an intermediate mixing magnitude. Below the RCB our eddy parameterization provides a stronger mixing due to the high T$_{int}$ assumed.

As for HAT-P-1b, we note that HCN presents a small mixing ratio in the upper atmosphere of this planet and peaks right below the homopause due to H$_2$O, N$_2$ and CO photodissociation (\cref{Fig:ChemEddy}).
The major part of the HCN photolysis mass flux therefore happens in this small region where the HCN mixing ratio is actually twice larger in the parameterized eddy case compared to the nominal, thus enhancing its contribution to the haze formation.
This larger HCN abundance is related to slightly larger CO and N$_2$ mixing ratios in this same region enhancing the local production of HCN.
However, the enhancement of HCN contribution to the haze formation is still negligible compared to the CO contribution (\cref{Tab:PhotoFluxEddy}). 
The latter remains the main haze precursor providing a three times larger haze mass flux in the parameterized eddy case compared to the nominal owing to larger abundances between 1 and 0.1 µbar.
This three times larger haze mass flux produces twice more numerous particles in the haze formation region (\cref{Fig:Haze}).
As particles settle down, the larger atmospheric mixing provided by our eddy parameterization hampers the particle coagulation compared to the nominal eddy case, further increasing the particle number density.
However, the differences remain weak and the haze mass flux small, thus resulting in negligible modification on the spectra (\cref{Fig:SpectraEddy}).
Nevertheless, we observe slightly weaker transit depths in the parameterized eddy case related to the weak haze opacity permitting to probe deep in the atmosphere.
Indeed, below 1 mbar, the haze optical depth is weaker in the parameterized eddy case compared to the nominal related to the smaller particle size, thus decreasing the transit depth.

\subsubsection{WASP-31b}

Above the 1 bar level, the difference between the parameterized and nominal eddy profiles remains smaller than an order of magnitude, while below, the parameterized eddy is larger compared to the nominal by up to four orders of magnitude from 10 to 100 bar (\cref{Fig:SelfEddies}). 
Despite these strong deep atmosphere differences, we have small modifications of the quench levels due to the hot temperatures producing high quench levels.
Like HAT-P-1b and HD-209458b, the major part of CO and HCN contribution to the haze formation happens around 1 µbar.
The larger CO mixing ratio, as well as, the resulting larger HCN abundances (\cref{Fig:ChemEddy}), enhance the haze production by a factor of 7.6 (\cref{Tab:PhotoFluxEddy}).
However, particle sublimation happens in the upper atmosphere of WASP-31b reducing the haze mass flux effectively produced.
Therefore, while the photolysis mass flux increases by a factor of $\sim$8, the effective haze mass flux increases by a factor of only 4 relative to the nominal eddy case.
Despite this 4$\times$ larger effective haze mass flux, the particle number density in the haze production region only increases by 40\% compared to the nominal eddy case (\cref{Fig:Haze}), owing to the stronger parameterized eddy magnitude in this region.
The weaker coagulation results in smaller and more numerous particles below the production region relative to the nominal eddy case.
However, such a small haze mass flux is not sufficient to have a significant impact on the spectrum of WASP-31b. 
Thus, the modifications brought by the parameterized eddy on the haze distribution do not affect the transit spectrum (\cref{Fig:SpectraEddy}) neither the temperature profile (\cref{Fig:TempEddy}) of this atmosphere.

\subsubsection{WASP-17b}

For WASP-17b, our eddy parameterization provides a profile close to the nominal in the region from 100 to 0.01 mbar where quenching occurs (\cref{Fig:SelfEddies}).
As a result, there are relatively small modifications of the quench level locations, hence on the haze precursors abundances.
Moreover, the large atmospheric temperatures demonstrated by WASP-17b make mixing rather inefficient and result in low abundances for CH$_4$, C$_2$H$_2$ and HCN in the upper atmosphere (\cref{Fig:ChemEddy}).
However, we note in the region from 1 to 0.1 µbar larger HCN and CO abundances by a factor of $\sim$2, similar to what is observed for HAT-P-1b and HD-209458b.
As previously stated, this region represents the major part of the contribution of these species to the haze formation, therefore we observe an enhancement of the haze production by a factor of 2 (\cref{Tab:PhotoFluxEddy}).
However, despite the strong photolysis of haze precursors, previous results suggest a clear atmosphere for WASP-17b \citep{Sing16,Sedaghati16}, resulting in very small upper limits on the haze formation yield \citep{Arfaux22}.
Since the chemical profiles of the most radiatively active species (mainly CO and H$_2$O) remain similar and haze is not affecting the atmosphere, we observe negligible modifications of the thermal structure and of the transit spectrum of this planet compared to the nominal eddy case (\cref{Fig:SpectraEddy}).

\section{DISCUSSION}
\label{Sec:Discussion}

\subsection{Effect of T$_{int}$ on the cloud formation}

\begin{figure}
\includegraphics[width=0.5\textwidth]{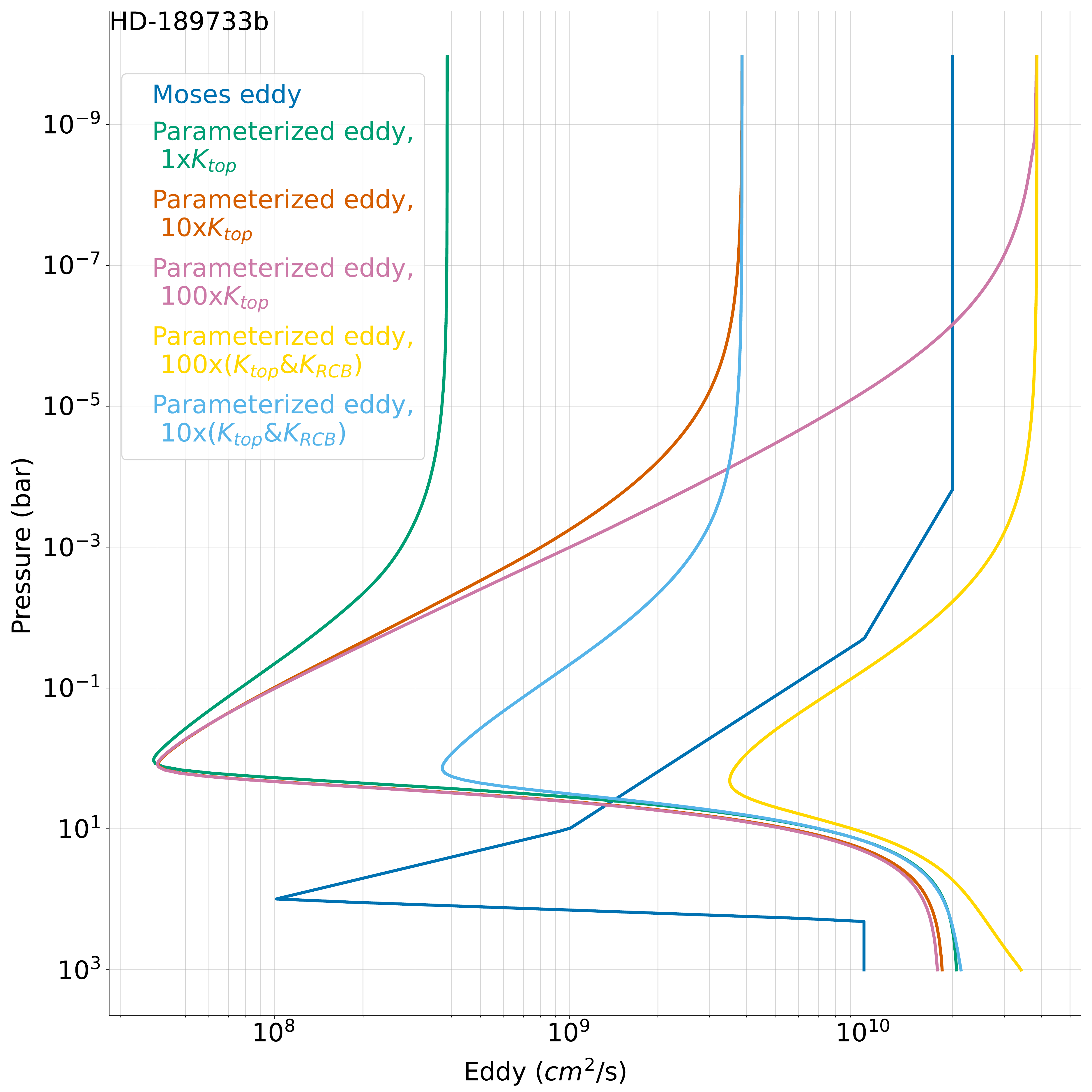}
\caption{Eddy profiles for HD-189733b with 10 and 100 times the homopause coefficient (orange and pink lines, respectively) and 10 and 100 times the homopause and RCB coefficients (yellow and cyan lines, respectively).
The blue is the \protect\cite{Moses11} profile and the green corresponds to the parameterized without scaling the homopause and RCB coefficients.
}
\label{Fig:189EDDY} 
\end{figure}

\begin{figure}
\includegraphics[width=0.5\textwidth]{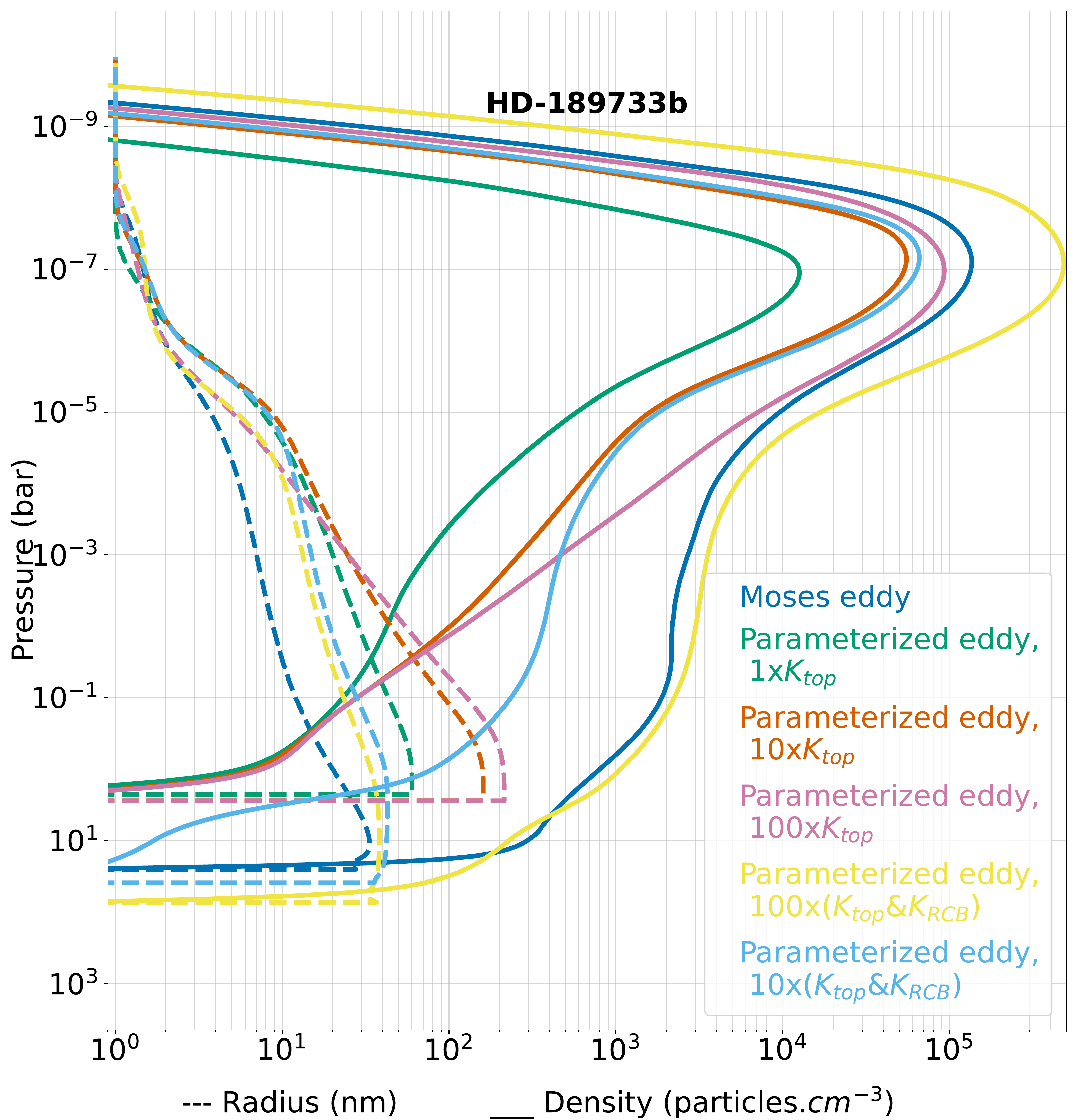}
\caption{Haze distribution for HD-189733b corresponding to the different cases tested in \protect\cref{Fig:189EDDY}.}
\label{Fig:189HAZE}
\end{figure}

\begin{figure}
\includegraphics[width=0.5\textwidth]{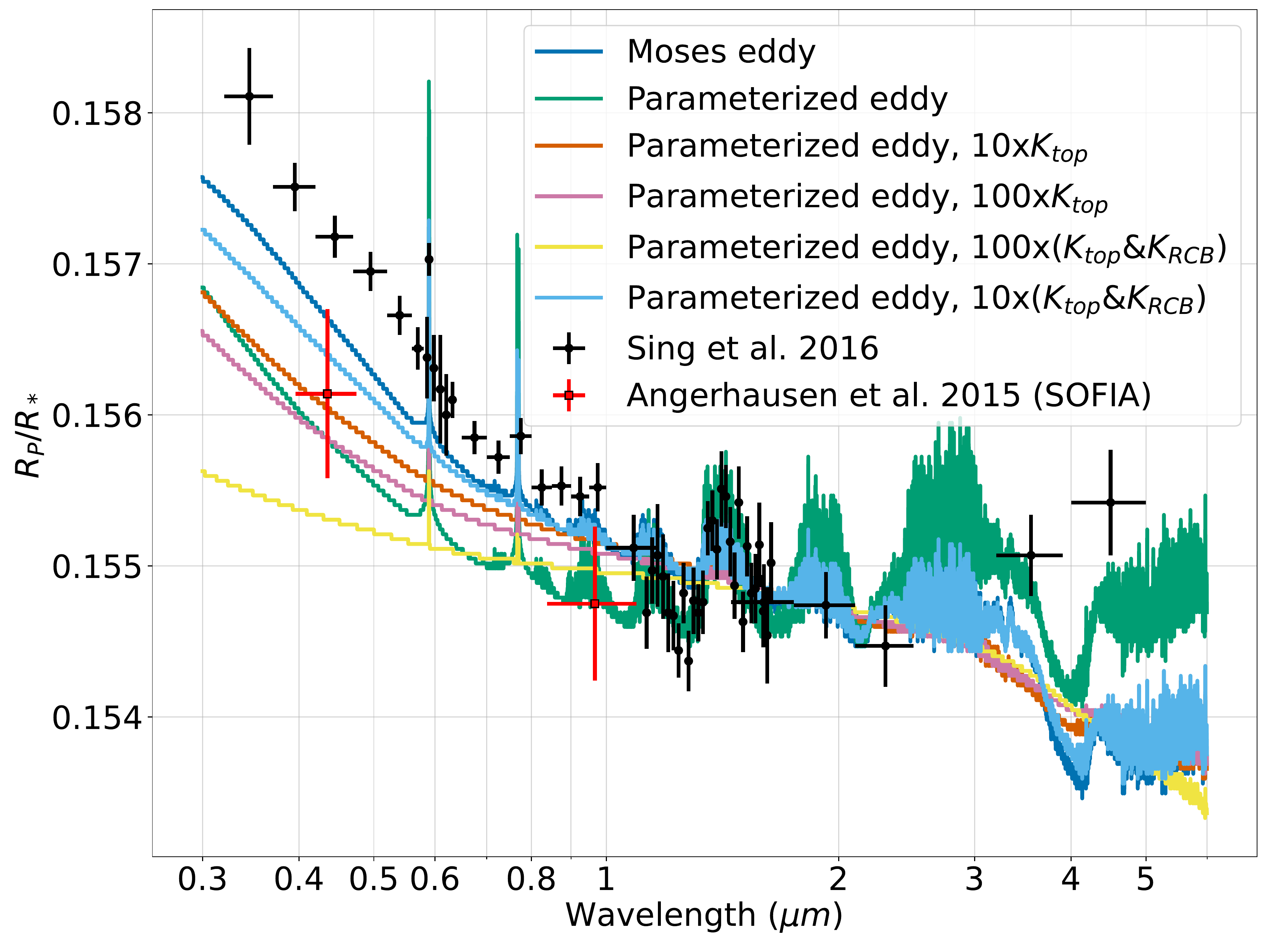}
\caption{Transit spectra for HD-189733b corresponding to the different cases tested in \protect\cref{Fig:189EDDY}.}
\label{Fig:189SPECTRA}
\end{figure}

Below the RCB, the atmospheric temperature follows the adiabatic lapse rate, hence decreases with increasing altitude.
With a hotter intrinsic temperature, the atmospheric temperature will follow an adiabatic lapse rate of higher entropy.
As the temperature profile crosses the condensation curve of a species, the latter is expected to condense and form cloud particles.
Therefore, higher T$_{int}$ values cause the cloud bases to shift towards higher altitudes making clouds more likely to impact the region probed by observations.

According to our simulations the iron cloud base rises from below the 1000 bar level in the T$_{int}$ = 100 K case to $\sim$100 bar in the higher intrinsic temperature cases of HAT-P-12b and WASP-6b, $\sim$10 bar for HAT-P-1b, HD-189733b, HD-209458b and WASP-39b, and $\sim$1 bar for WASP-17b and WASP-31b (\cref{Fig:TempEddy}).
For all planets except WASP-31b, the silicate formation follows a similar behavior of that of iron clouds with a cloud base below 100 bar in the T$_{int}$ = 100 K case and formation pressures similar to those of the iron cloud base in the high T$_{int}$ case.
In the specific case of WASP-31b, silicate clouds would already form high up in the atmosphere around 0.1 bar in the T$_{int}$ = 100 K case, and therefore, their formation is not impacted by the change of intrinsic temperature.
However, we note a small formation region for Mg$_2$SiO$_4$ near 1000 bar in the T$_{int}$ = 100 K case that could have cold-trapped the necessary material, thus hampering their formation higher up.
Sulphur clouds form at colder temperatures, therefore, for planets in which such clouds can appear, their formation region is higher up compared to iron and silicate clouds.
Thus, the sulphur cloud base for HAT-P-1b, HAT-P-12b and WASP-17b is weakly impacted by the change of intrinsic temperature.
However, for HD-189733b, WASP-6b and WASP-39b, MnS clouds could form below 10 bars in the T$_{int}$ = 100 K case, which could cold trap the necessary material preventing their formation higher up.
With the larger intrinsic temperature, this deep production region vanishes in the three planet cases, thus allowing the production of such clouds at much smaller pressures, above the 1 bar level for WASP-6b and WASP-39b and above the 0.1 bar level for HD-189733b.

Our cloud formation estimates are based on our full redistribution temperature profiles.
However, inhomogeneities between the morning and evening terminator conditions may further modify on the cloud composition and distribution \citep{Tsai22a,Feinstein22}.
Such effects need to be further explored in more detailed studies combining GCMs and cloud microphysics models.

\subsection{Homopause and RCB eddy coefficients}

For HD-189733b, the parameterization that we developed provides eddy profiles similar in magnitude to tracer-distribution-based GCM calculations in the middle atmosphere (between 1 bar and 1 mbar, \cref{Fig:SelfEddies}).
However, concerning the homopause eddy coefficient of HD-189733b, as well as, that of HD-209458b and WASP-31b, we notice a divergence between the \cite{Koskinen10} scaling and the GCM results.
Indeed, both passive tracers and root mean square vertical velocity methods suggest homopause values $\sim100\times$ larger than those obtained by scaling Jupiter's eddy coefficient.
On the other hand, this scaling of the homopause requires knowledge about the characteristic turbulent velocities that are not available for most planets.
Therefore, we used for all planets the value estimated by \cite{Koskinen10} for HD-209458b.
The use of a characteristic turbulent velocity factor tailored for each planet could provide better aggreement with the GCM estimates.

We test this hypothesis by multiplying the HD-189733b homopause eddy coefficient by factors of 10 and 100 to simulate a larger turbulent velocity.
This results in a higher haze production of 1.4x10$^{-12}$ and 4.0x10$^{-12} g.cm^{-2}.s^{-1}$ for the 10 and 100 x$K_{top}$ cases, respectively, against 6.4x10$^{-14} g.cm^{-2}.s^{-1}$ for the 1x$K_{top}$ case.
We however note that these mass fluxes remain lower than that obtained for the nominal eddy case of this planet. This happens because the eddy diffusion in the middle atmosphere remains small among the three parameterized eddy cases (green, orange and pink lines in \cref{Fig:189EDDY} for the 1, 10 and 100 xK$_{top}$ cases, respectively) compared to the nominal eddy profile (blue line in \cref{Fig:189EDDY}) and lead to the formation of larger particles (green, orange and pink lines in \cref{Fig:189HAZE} for the 1, 10 and 100 xK$_{top}$ cases, respectively) compared to the nominal eddy case (blue line \cref{Fig:189HAZE}). Furthermore, these parameterized profiles cause a decrease in the steepness of the UV-visible slope of the simulated transit spectra (\cref{Fig:189SPECTRA}) and cannot reproduce the steep UV-visible slope obtain with the nominal eddy profile, let alone fit the HST observations. 

A solution appears to be a joint increase of both the homopause (K$_{top}$) and RCB (K$_0$) eddy coefficients.
We therefore evaluate the parameterized eddy where we multiply these parameters by factors of 10 and 100 (respectively cyan and yellow lines in \cref{Fig:189EDDY,Fig:189HAZE,Fig:189SPECTRA}).
The latter case provides a large haze mass flux of 1.3x10$^{-10} g.cm^{-2}.s^{-1}$ much beyond the value obtained with the nominal eddy profile (9.0x10$^{-12} g.cm^{-2}.s^{-1}$).
Although this 100x$K_{top}\&$100x$K_0$ case produces an eddy profile slightly larger than the nominal profile, the large haze mass flux results in large and numerous particles with a grey behavior and therefore a rather flat transit spectrum (\cref{Fig:189SPECTRA}).
In the 10x$K_{top}\&$10x$K_0$ case however, though it provides eddy diffusion values $\sim$20$\times$ smaller than the nominal profile, the weaker haze formation of 2.0x10$^{-12} g.cm^{-2}.s^{-1}$ reduces the number of particles compared to the 100x$K_{top}\&$100x$K_0$ case, and results in a steep UV-visible slope that matches well the HST/STIS data.
Our parameterization, therefore, can be tailored for each planet by consistently tuning the RCB and homopause eddy coefficients.
Nevertheless, we note again that our standard parameterization (1x$K_{top}\&$1x$K_0$) provides a good match of the SOFIA observations of HD-189733b \citep{Angerhausen15}, while we expect the HST-STIS observations to be more affected by stellar activity \citep{McCullough14}.

\section{CONCLUSIONS}
\label{Sec:Conclusions}

In this study, we developed a parameterization for the eddy diffusion profile based on convection in the deep atmosphere, gravity waves break down in the middle atmosphere and turbulence in the upper atmosphere.
Our parameterization is impacted by the planetary intrinsic temperature (T$_{int}$), we therefore also studied the impact of increasing the intrinsic temperature using the values suggested by \cite{Thorngren19b} compared to a T$_{int}$ = 100 K case.
We then proceeded to evaluating the implications of the high T$_{int}$ on the eddy profile and the ramification of the latter on the atmospheric composition and haze properties.

Our results on increasing the intrinsic temperature demonstrates very little impact in the region of the atmosphere probed by the observations, and negligible changes for the temperature profile, main chemical species and haze distributions in the upper atmosphere.
Indeed, the main modifications of the major atmospheric constituent are confined to the deep atmosphere where thermochemical equilibrium dominates and the temperature profiles from both high and low T$_{int}$ cases converge before transport starts to have an impact on major species.
However, for equilibrium temperatures lower than 1300 K, we find that the haze precursors are impacted by the changes arising in the deep atmosphere.
We especially note lower HCN abundances in the upper atmosphere (related to lower ammonia quenching mixing ratio in the deep atmosphere) compared to the T$_{int}$ = 100 K case, which result in a weaker haze formation.
However, this effect is counterbalanced when the implications of the high T$_{int}$ are accounted for in the eddy profile.

The proposed eddy parameterization can be easily adapted to other studies since it depends on only two free parameters, the eddy magnitude at the radiative convective boundary (K$_0$) and its corresponding magnitude at the homopause (K$_{top}$).
Using common values for K$_{0}$ and K$_{top}$ among the studied planets, we derive mixing profiles of similar shape and magnitude as the nominal profile used previously, though slightly larger within one order of magnitude above the radiative-convective boundary.
Below the radiative convective boundary, the parameterized eddy is significantly larger  than what is proposed in previous studies, due to the high T$_{int}$ values of the studied planets. 
For haze-free planets, the parameterized eddy results in negligible changes for the main atmospheric species and the thermal structure.
However, for hazy planets, multiple phenomena happen resulting in changes in the transit spectra.
First, the more efficient redistribution of HCN enhances its upper atmosphere abundance, thus increasing its contribution to haze formation.
As the haze formation yield is kept fixed in this set of simulations, we obtain enhanced haze mass fluxes compared to the nominal eddy case.
In addition, the stronger eddy profile hampers the particle coagulation and then leads to smaller but more numerous particles, the latter being further enhanced by the larger haze mass flux.
These smaller and more numerous particles result in a steeper UV-visible slope of the transit spectra, as well as, larger transit depths.

Although using common K$_{0}$ and K$_{top}$ values provided suitable fits of the transit observations for most of the studied planets, we note the particular case of HD-189733b that requires a much stronger eddy profile to fit the HST observations.
In this specific case, our eddy parameterization shows its limitations indicating that further knowledge is required to constrain the homopause and radiative-convective boundary eddy diffusion coefficients.
Our simulations suggest that these two parameters can be consistently adjusted by a common scale factor ($\sim$10), relative to the nominal values assumed, to provide an eddy mixing profile consistent with estimates from GCMs. 
This atmosphere also demonstrates a particular behavior when increasing the intrinsic temperature, with HCN contribution dropping and C$_2$H$_2$ becoming a main haze precursor, contrary to the T$_{int}$=100 K case.
Both cold temperatures and strong eddy mixing are playing the role of shifting the quench levels at higher pressures making them sensitive to modifications of the intrinsic temperature.
Therefore, cold and/or strongly mixed exoplanet atmospheres are sensitive to changes in the intrinsic temperature while hotter planets present very weak modifications.

Finally for the case of WASP-39b, recently studied with JWST, we find that the higher metallicity inferred for this atmosphere necessitates the presence of clouds/hazes contrary to previous estimates based on the assumption of solar metallicity. 
We further find that both cloud and hazes seem to be necessary for satisfying both the JWST and the HST observations.

\section*{Acknowledgements}

We thank our reviewer for fruitful comments and suggestions on the manuscript. We acknowledge support from the Programme National de Planétologie (PNP) of INSU/CNRS through the project TISSAGE.

\section*{Data Availability}

The data underlying this article will be shared on reasonable request to the corresponding author.



\bibliographystyle{mnras}
\bibliography{biblio} 





%


\bsp	
\label{lastpage}
\end{document}